\newif\iflongversion
\longversionfalse
\iflongversion
\documentstyle[prb,preprint,aps]{revtex}
\else
\documentstyle[prb,aps,multicol,epsfig]{revtex}
\fi

\begin{document}
\title{Half-filled Landau level as a Fermi liquid of dipolar quasiparticles}

\author{Ady Stern,$^1$, Bertrand I. Halperin$^2$, Felix von
  Oppen$^{3,4,}$\cite{felix}, Steven H. Simon$^5$} 

\address{$^1$ Department of Condensed
  Matter Physics, The Weizmann Institute of Science, Rehovot 76100, Israel\\
  $^2$ Physics Department, Harvard University, Cambridge,
  Massachusetts 02138 \\
  $^3$ Max-Planck-Institut f\"ur Kernphysik, Postfach 103980, 69029
  Heidelberg, Germany\\
  $^4$ Institut f\"ur Theoretische Physik, Universit\"at zu K\"oln, 
  Z\"ulpicher Str.\ 77, 50937 K\"oln, Germany\\
  $^5$ Lucent Technologies Bell Labs, Murray Hill, NJ 07974}
  
\date{\today}

\maketitle
\def\kf{k_{\mbox{\tiny{F}}}}  

\begin{abstract} 
  In this paper we study the relation between the conventional
  Fermion-Chern-Simons (FCS) theory of the half-filled Landau level
  ($\nu=1/2$), and alternate descriptions that are based on the notion
  of neutral quasi-particles that carry electric dipole moments.  We
  have previously argued that these two approaches are equivalent, and
  that e.g., the finite compressibility obtained in the FCS approach
  is also obtained from the alternate approach, provided that one
  properly takes into account a peculiar symmetry of the dipolar
  quasiparticles --- the invariance of their energy to a shift of
  their center of mass momentum.  Here, we demonstrate the equivalence
  of these two approaches in detail.  We first study a model where the
  charge and flux of each fermion is smeared over a radius $Q^{-1}$
  where results can be calculated to leading order in the small
  parameter $Q/\kf$.  We study two dipolar-quasiparticle descriptions
  of the $\nu=1/2$ state in the small-$Q$ model and confirm that they
  yield the same density response function as in the FCS approach.  We
  also study the single-particle Green's function and the effective
  mass, for one form of dipolar quasiparticles, and find the effective
  mass to be infra-red divergent, exactly as in the FCS approach.
  Finally, we propose a form for a Fermi-liquid theory for the dipolar
  quasiparticles, which should be valid in the physical case where $Q$
  is infinite.
\end{abstract}

\def\hatn#1{{\bf{\hat{#1}}}}
\def\kf{k_{\mbox{\tiny{F}}}}              
\def\vf{v_{\mbox{\tiny{F}}}}
\def\rd{\rho_{\mbox{\tiny{D}}}}            
\def\cf{{\mbox{\tiny{CF}}}}
\def\saw{{\mbox{\tiny  SAW}}} 
\def\cs{{\mbox{\tiny CS}}}
\def\vec#1{{\bf #1}}
\def\newpsi{\psi^{\phantom{\dagger}}}
\def\newpsidagger{\psi^{\dagger}}
\def\mycaption#1{\scriptsize #1} 
\def\myscriptsize#1{\mbox{\scriptsize #1}}
\def\mathscriptsize#1{\mbox{\scriptsize $#1$}}
\def\Kfr{K^{\mbox{\tiny{free}}}}

\iflongversion
\typeout{Skipping Multicols}
\else
\begin{multicols}{2}
\fi


\section{Introduction}
\subsection{Overview}

Many aspects of the behavior of an interacting two-dimensional
electron system in the lowest Landau level have been understood using
the ``composite fermion'' picture\cite{CFgeneral,sarma}.  This description,
in practice, includes a number of related concepts and computational
methods.  The term ``composite fermion'' was first introduced by
Jain\cite{Jain} in connection with the trial wave functions that he
used to describe electronic ground states at the most prominent
quantized Hall plateaus, where the filling fractions are of the form
$\nu=p/(2p+1)$, with $p$ an integer.  A ``Fermion-Chern-Simons'' (FCS)
approach was employed by Lopez and Fradkin\cite{Lopez}, and
others\cite{Others} as an alternate way of understanding the composite
fermion ground states and as a method for understanding the spectrum
of collective excitations in the quantized Hall states. The FCS
approach was also used by Halperin, Lee and Read\cite{HLR} (HLR) and by
Kalmeyer and Zhang\cite{KZ} to describe
phenomena at, or near, even-denominator filling fractions, such as
$\nu = 1/2$, where quantized Hall plateaus are not observed.  The HLR
analysis predicted that the ground state at these filling fractions
should be {\it compressible}\cite{compdef}, 
and that its properties could be understood by
a perturbative analysis, starting from a ground state which is a
filled Fermi sea of appropriately defined fermions.

The FCS approach begins with a unitary transformation, in which the
electron system is converted to a system of fermions that interact
with each other via a fictitious gauge field ${\bf a_{\cs}}$ of the
Chern-Simons type, as well as via the usual electromagnetic Coulomb
repulsion. The Hamiltonian of the fermions is given below in 
Eq. (\ref{ham}).

A key prediction of the FCS theory is that the quasiparticles at a
filling fraction such as $\nu=1/2$ can travel in straight lines over
large distances, oblivious to the effects of the strong applied
magnetic field.  Within the FCS formulation used by HLR\cite{HLR},
this is most naturally explained by saying that the quasiparticles
feel an effective magnetic field $\Delta B$, which is the the
difference of the applied magnetic field and the mean value of a
fictitious Chern-Simons magnetic field, and these two contributions
cancel each other precisely at $\nu=1/2$. At filling fractions
slightly away from $\nu=1/2$, the cancellation is not perfect, so
$\Delta B \ne 0$.  As a result, the quasiparticles should move in a
circle, with a radius given by the ``effective cyclotron radius''
$R^*_c = \kf/|\Delta B|$, a prediction which has been confirmed by
several experiments\cite{Experiments}. (Here $\kf$ is the
Fermi-momentum, related to the electron density $n_0$ by $\kf = (4 \pi
n_0 )^{1/2}$.  We use units where $\hbar = 1$, the speed of light is
$1$, and the electron charge is -1.)

An alternate view of the HLR predictions, which was emphasized by
Read\cite{NickRead}, is to say that the actual low-energy
quasiparticles, obtained from the bare fermions of the HLR theory
after ``screening'' by relaxation of the high-frequency magnetoplasma
modes (or Kohn modes) which occur at the bare cyclotron frequency of
the electron system, are actually electrically neutral at $\nu=1/2$.
At nearby filling fractions, the low energy quasiparticles have charge
$e^* = (2 \nu - 1)$, which coincides with the quasiparticle charge
$-1/(2p+1)$ expected for the fractional quantized Hall states at
filling fractions $\nu = p/(2p+1)$.\cite{Endnote1}  A quasiparticle
with charge $e^*$ which sees the full magnetic field $B$ will have the
same effective cyclotron radius $R^*_c$ as a fermion of charge -1 that
sees the effective field $\Delta B$.  Another important point which
was also noted by Read, is that the low energy quasiparticles at
$\nu=1/2$, though they are overall neutral, carry an electric dipole
moment which is proportional to, and perpendicular to, the canonical
momentum of the quasiparticle\cite{NickRead,SimonReview}.
 
Although the FCS theory has had many successes, and has several
advantages over other formulations, it also has several disadvantages.
First, the coupling to the Chern-Simons field, which is one of the
parameters of the perturbation expansion, is not small in any case of
physical interest.  Second, the FCS analysis does not directly reveal a
key feature of the system of electrons in the lowest Landau level:
namely that all the intra-Landau-level excitation-energies must vanish
in the limit where the Coulomb repulsion is taken to zero. (An
equivalent statement is that the effective masses and excitation
energies remain finite if the band mass tends to zero while the
strength of the electron-electron repulsion is held fixed.)  Another
disadvantage of the FCS approach is that it employs a perturbation
theory where the fundamental entities are bare fermions with
charge -1, which are quite far from the actual low-energy quasiparticles.
(The true charge of the quasi-particles is revealed only
after their interaction with the gauge field is taken into account, as
described above.)

Various alternate formulations of the composite fermion picture, which
attempt to overcome the disadvantages explained above, have been
recently proposed\cite{Shankar,Pasquier,ReadNew,DHLee}. Several
authors have proposed formulations of composite-fermion theory which,
by construction, lie entirely within the subspace of the lowest Landau
level\cite{Pasquier,ReadNew,DHLee}. These formulations automatically
incorporate the feature that the energy scale and effective mass are
set by the electron-electron interaction. (A recent formulation by D-H
Lee\cite{DHLee} also incorporates the feature of exact particle-hole
symmetry at $\nu=1/2$ for electrons restricted to the lowest Landau
level).

In the present paper, however, we shall be more interested in an
approach proposed recently by Murthy and Shankar\cite{Shankar} (MS).
MS begin with the exact Hamiltonian of the FCS approach, given by Eq.
(\ref{ham}) below, and make a unitary transformation in which the 
field operators for the fermions and the Chern-Simons vector potential
in the FCS theory, $\newpsidagger_\cs,\newpsi_\cs,{\bf a}_\cs$, are transformed
to a new set of operators $\newpsidagger,\newpsi,{\bf a}$, which
have several desirable features. First, in the transformed
Hamiltonian the fermionic operators $\newpsidagger,\newpsi$ are approximately
decoupled from the gauge field  ${\bf a}$. Second, the transformed
fermions do indeed carry electric dipoles at $\nu=1/2$. And third, by
treating the transformed Hamiltonian in an appropriate approximation,
Murthy and Shankar obtain results for the effective mass which are
tied to the strength of the electron-electron repulsion and
independent of the bare mass.

If the low energy quasiparticles are decoupled from the high-energy
magnetoplasma modes at long wavelengths, it
should be possible to describe the low-frequency response of the
electron system at $\nu=1/2$ directly in terms of these quasiparticles.
However, an attempt to do this leads to a paradox, first noted by
Shankar and Murthy\cite{Shankar}: if the dipolar fermions are assumed
to behave like a conventional Fermi liquid of neutral particles which
carry only an electric dipole moment, then one finds that it is
costly, in terms of kinetic energy, to produce a fluctuation in the
charge density at long wavelengths, and one finds that the electron
system is incompressible, i.e., that the zero-frequency electron
density response function at $\nu=1/2$ vanishes $\propto q^2$ in the
limit $q\rightarrow 0$, which would be contrary to the predictions of HLR.
The possibility of incompressibility at $\nu =
1/2$ was also suggested in preprints by several other authors
\cite{Pasquier,DHLee}.

The resolution of this paradox was outlined in a Comment\cite{Comment}
by two of the present authors (HS). As had been noted earlier by
Haldane\cite{Haldane1} for the case of composite fermions restricted
to the lowest Landau level, the system of dipolar fermions has the
special property that the total energy is unchanged if a constant
${\bf K}$ is added to the momentum of every particle.  (This property,
which we call ``total-momentum invariance'', or $K$-invariance, does
not hold, of course, for free fermions).  Consequently, as noted by
HS, it costs very little energy to produce a long wavelength
fluctuation in the transverse momentum density of the dipolar
particles.  Since the relation between the quasiparticle dipole moment
and its momentum implies that the electron charge density is
proportional to the gradient of the transverse momentum density, HS
argued that the $K$--invariance implies the finite compressibility of
the electron system.  The possibility of such a resolution of the
paradox was mentioned in the original publication of Shankar and
Murthy\cite{Shankar}, who speculated that the ``drifting Fermi sea''
peculiar to this system might lead to a finite compressibility.  The
conclusion of finite compressibility was also reached more recently by
Read\cite{ReadNew} and by D-H Lee (in a revised version of Ref.
\onlinecite{DHLee}), independently of HS.

It must be noted that the electronic density--density linear response
function is a measurable quantity, which should determine the outcome
of experiments measuring surface acoustic wave propagation,
compressibility, magneto-capacitance, Coulomb drag, and optical
response\cite{Experiments}. Results of all of these experiments have
been interpreted in terms of the predictions of the conventional FCS
theory, and would seem to support the prediction of finite
compressibility\cite{Experiments}.  Nevertheless, it is difficult to
interpret them as definitively ruling out the possibility of
incompressibility at $\nu=1/2$, and MS have argued that some of the
results can be fit quite well using a response function which has the
feature of incompressibility.  The experimental situation is
complicated further by the presence of impurities, which might give
rise to a finite compressibility even if the pure system were
incompressible.  Thus it is comforting that the issue can be resolved
on theoretical grounds, and that one finds the pure system to be
compressible within all the current approaches.  

The purpose of the present paper is to explore in greater detail the
resolution of the compressibility paradox. We wish to see not only how
the dipolar gas leads to a finite compressibility, but to examine more
closely the relation between the FCS approach and several possible
formulations in terms of dipolar fermions.  We begin by studying a
modification of the model {\it where it is possible to calculate the static
compressibility and the dynamic response functions exactly, to lowest
order in a small parameter}, using either the FCS approach or an
approach based on dipolar quasiparticles. We verify that in that
calculation, one obtains exactly the same density-density response
functions using the different approaches. 
 
\subsection{The Fermion-Chern-Simons Approach}

The unitary transformation, which transforms the electron system to a 
composite fermion description in the FCS approach, 
was originally
employed by Leinaas and Myrheim\cite{Leinaas} in the 1977 paper which
introduced the concept of fractional statistics for two-dimensional
systems.  We shall follow common usage, however, and refer to the
transformation as a ``Chern-Simons transformation,'' and we shall refer to
the resulting transformed fermions as ``(bare) Chern-Simons fermions'',
or ``CS fermions.''

For $\nu=1/2$, in the temporal (Weyl) gauge for the Chern-Simons gauge
field ($a_0=0$), the Hamiltonian takes the form
%
%
\begin{eqnarray}   \label{ham}
& & H_\cs =
\int d\vec r\frac{1}{2m} \left|(-i{\bf \nabla -a_{\cs}})\newpsi_{\cs}({\bf
  r})\right|^2 \\  +
&\frac{1}{2}&
\int d\vec r d\vec {r'} \, \left(|\newpsi_{\cs}({\bf r})|^2-n_0 \right) \, 
v({\bf r}-{\bf r'})
\, \left(|\newpsi_{\cs}({\bf r'})|^2-n_0 \right) \nonumber
\end{eqnarray}
Here the two components of the Chern--Simons vector potential ${\bf
  a}_{\cs}$ are canonically conjugate, satisfying 
$[a_x({\bf r}),a_y({\bf r'})]=
i2\pi{\tilde\phi}\delta({\bf r}-{\bf r'})$, $n_0$ is the average
electron density, $v(\vec r)$ is the electron-electron interaction, and the
even integer $\tilde\phi$ is the number of flux quanta attached to
each electron by the Chern--Simons transformation. (For $\nu=1/2$, we choose
$\tilde\phi=2$.)  
The vector potential due to the applied magnetic field has been 
absorbed by shifting the Chern-Simons field ${\bf a_{\cs}}$.
Hence, in the Weyl gauge, the Chern-Simons field 
is dynamic and we
have enlarged the Hilbert space compared to the Coulomb gauge used in HLR. The
physical states of the theory are those which satisfy the Chern-Simons
constraint,
\begin{equation}
{\bf \nabla}\times {\bf a}_\cs=2\pi{\tilde\phi}(\newpsidagger_\cs\newpsi_\cs
-n_0).  
\label{csconstraint}
\end{equation}
The density operator for the bare CS fermions, $\newpsidagger_\cs\newpsi_\cs$,
is the same in the Weyl and Coulomb gauges, and is identical to the density
operator for the electrons.

In the FCS approach the ground state of the Hamiltonian (\ref{ham})
and the energies of low-lying excited states are first obtained in the
Hartree approximation; then one calculates response functions and
corrections to the excitation energies using the random phase
approximation\cite{Lopez,HLR} (RPA) or more sophisticated
approximations based on a Feynman-diagram analysis of the
perturbations arising from the Chern-Simons and Coulomb
interactions\cite{SternHalperin,DivergencesCancel,Marston,Aim}.  In
particular, these calculations conclude that the $\nu=1/2$ state is
``compressible''; i.e., the zero frequency density-density response
function of the electrons is found to be finite in the limit of small
wavevector $q$ if the interaction between the electrons is of short
range, and to equal $\frac{q\epsilon}{2\pi e^2}$ if the electrons
interact via an unscreened Coulomb interaction (here $\epsilon$ is the
dielectric constant). Furthermore, it is concluded that the relaxation
of charge fluctuations follows a dispersion law of $\omega\propto i
q^3$ for short range electron-electron interaction (and $\omega\propto
iq^2$ for unscreened Coulomb interaction). In addition, it has been
argued that the effective mass of the quasi-particles should diverge
close to the Fermi surface (logarithmically in the case of Coulomb
interactions), and that this divergence should be manifest in the
behavior of the energy gaps of the quantized Hall states close to
$\nu=1/2$.

The modified model used for our calculations employs a momentum cutoff
$Q \ll \kf$ for the Chern-Simons interaction.  A similar cutoff was
used previously by one of the authors\cite{HalperinAnyon} in the
context of anyon models of high-temperature superconductivity.  More
recently, such a cutoff was employed within the context of the quantum
Hall effects by Raghav Chari {\it et al.}\cite{HaldaneSmear}, who
termed the approach ``fat flux quanta''.  The modified model, which is
defined more precisely below (see Eq. \ref{hamq}), may be thought of
as a system of composite fermions whose charge and flux tubes are
spread over a finite radius, of order $Q^{-1}$. Barring singular phase
transitions, the use of a small value of $Q$ should make the RPA
exact, to lowest order in $Q$, and it enables us to carry out
explicitly the Murthy-Shankar unitary transformation from CS fermions
to dipole fermions. The Hamiltonian we obtain for the dipole fermions
is not that of free dipolar particles, but rather a more complicated
one, given by Eq.  (\ref{hd}) below. The Hamiltonian contains several
terms which are additional to the ones considered explicitly by MS.

\subsection{Alternative Definitions of the Quasiparticles}

An important point, which will be discussed further below, is that in
intermediate stages, various definitions of the low energy dipolar
quasiparticles are possible, which are not identical in the long
wavelength limit, even though they all lead to the same 
density-density response function for the electrons.  
In all of the descriptions we
consider here, for an infinitesimal fluctuation about $\nu = 1/2$, the
electron density $\rho^e$ is given by the relation
\begin{equation}
\rho^e = \frac{\nabla \times {\bf g}}{2 \pi {\tilde \phi} n_0} 
\label{rhoe}
\end{equation}
where {\bf g} is the momentum density of the quasiparticles. 
Eq. (\ref{rhoe}) relates the momentum density to the charge 
density.  Nevertheless, one has freedom in the precise
definition of the position of a quasiparticle, and hence in the
relation between the electron density and the density of
quasiparticles at non-zero wavevector ${\bf q}$.

In particular, in Section II below we shall derive a description of the
half-filled Landau level that employs dipolar quasiparticles which
obey a constraint
\begin{equation}
\rho({\bf q})=
\frac{i{\bf
    q}\times{\bf g}}{2\pi{\tilde \phi}n_0}
\label{constraint2}
\end{equation}
and whose density is related to the electron density by 
\begin{equation}
\rho({\bf q}) =  \rho^e({\bf q}) 
\label{rhorhoe}
\end{equation}
One sees that in the long-wavelength limit the positions of the quasiparticles 
employed in Section II must coincide, on average, with the positions of the 
electrons in the system.  We shall describe such quasiparticles as 
``electron-centered'' quasiparticles.

In section III below we discuss 
the quasiparticles used by MS, which obey a constraint 
\begin{equation}
\rho({\bf q})=\frac{i}{2}
\frac{{\bf
    q}\times{\bf g}}{2\pi{\tilde \phi}n_0}
\label{constraint}
\end{equation}
while the relation between $\rho^e $ and the quasiparticle density
$\rho$, for $q \ne 0$, is $ \rho({\bf q}) = (1/2) \rho^e({\bf q}) $.
The positions of these quasiparticles are shifted, on average,
relative to the electron positions, by an amount $ {\hat z} \times
<{\bf g}({\bf r})> / (4 \pi {\tilde \phi} ) $ where $<{\bf g} ( {\bf
  r})>$ is the momentum density averaged over a volume of radius
$Q^{-1}$ about the position of the quasiparticle.  The relations
(\ref{rhoe}) and (\ref{constraint}) would be satisfied by a collection
of dipolar quasi-particles when the dipole moment \ of a
quasi-particle of momentum $\bf k$ is $-\frac{{\hat z}\times {\bf
    k}}{2\pi{\tilde\phi}n_0}$ and {\it the position of the
  quasi-particle is defined to be half-way between the negative and
  positive charges of the electric dipole.} We describe quasiparticles
of this type as ``shifted quasiparticles''.

Although the explicit calculation we carry out is to lowest order in
$Q$, we believe that it sheds light also on the physical case, in
which no upper cut-off to $Q$ is present. In particular, it leads us
to show that while the density response of a ``conventional'' Fermi
liquid of dipoles to a scalar potential would be weak, leading to an
incompressibility of the liquid, this is not true once the fermions
satisfy $K$-invariance and their dipole moment is proportional and
perpendicular to their momentum.  Under these conditions, the fermions
density-density response function has the $q,\omega$ dependence one
obtains from the FCS approach (for small $q$ and $\omega$)
independent of whether $Q/\kf$ is small or large.

\subsection{Outline}

The outline of the paper is as follows. In the next section, we show
how one can derive a description in terms of electron-centered dipolar
quasiparticles, in a simple fashion, from the fermion Chern-Simons
Hamiltonian (\ref{ham}).  This is done most naturally by using a
Lagrangian formalism, and eliminating the high energy modes, as well
as the Chern-Simons vector potential, in an RPA-like approximation to
the Lagrangian.  In Section III, we obtain shifted quasiparticles, from
the same starting Hamiltonian (\ref{ham}), using a Hamiltonian
formalism and the unitary transformation employed by MS, for the model
with small momentum-cutoff $Q$.  In both cases, we calculate the
electron density response function, and obtain the same results as
originally obtained by HLR.  In Section IV, we make some additional comments
on the physical significance of the small-$Q$ model and the validity of the
RPA in that model.  In Section V, we discuss the effective mass and
the one-fermion Green's function
for the shifted quasiparticles employed in Section III, in the small-$Q$
model, and compare them to results of the conventional FCS approach.
The Fermi-liquid description which we believe to be valid in the actual
$\nu=1/2$ system, where there is no small cutoff $Q$, is discussed in Section
VI, for the case of electron-centered quasiparticles.  Our conclusions are
then summarized in Section VII.  Some details of the response matrix used
in the calculations of Section III are presented in an Appendix.

\section{Electron-centered quasiparticles: a Lagrangian approach }
\label{sec:Lagrange}

In this section, we begin with a Lagrangian formulation equivalent to
the Hamiltonian (\ref{ham}) of the FCS theory in the temporal gauge, and
integrate out the variables associated with the Chern-Simons
field. Using the Random Phase Approximation, we obtain the same
density response function as was obtained by HLR, using the Coulomb
gauge for the Chern-Simons field.  We show, however, that the low
energy part of the fermionic theory has a natural interpretation in
terms of dipolar quasiparticles whose density coincides with the
density of the electrons, at long wavelengths (i.e., they are
``electron-centered'' quasiparticles.)  We also find an explicit form
for the effective Lagrangian of the quasiparticles at low energies,
using approximations that, along with the RPA, are valid in the limit
where there is a wavevector cutoff $Q$ which is small compared to
$\kf$.

The zero temperature action corresponding to the Hamiltonian (\ref{ham}) is
\begin{eqnarray}
 \label{Weyl-action}
   S=\int dt\,d{\bf r}&&\left\{{1\over2}{1\over2\pi\tilde\phi}
   (a_x\partial_0 a_y-a_y\partial_0
   a_x)+\bar\newpsi i\partial_0\newpsi\right.
   \nonumber\\
   &&\left.-{1\over2m}\bar\newpsi(-i\nabla+{\bf A}-
   {\bf a})^2\newpsi\right\}+S_{\rm Coul}.
\end{eqnarray}
We drop the subscript CS in this section, because we will not
consider here a canonical transformation of the fields.  Using the constraint
(\ref{csconstraint}) inherent in the CS formulation for $\nu=1/2$, the
Coulomb interaction can be written in terms of the CS field as
\begin{eqnarray}
\label{Coulomb-interaction}
  S_{\rm Coul}&&=-{1\over2}{1\over(2\pi\tilde\phi)^2}
   \nonumber\\
   &&\times
   \int dt\,d{\bf r}d{\bf r^\prime}[\nabla\times{\bf a}({\bf r})]
   v({\bf r}-{\bf r^\prime})[\nabla^\prime\times{\bf a}({\bf r^\prime})].
\end{eqnarray}
We included in the action an external vector potential ${\bf A}$ as a
source field for generating the electronic current response function
$K^e_{\alpha\beta}({\bf q},\omega)$. For the purposes of this section,
it will be useful to consider $K^e_{\alpha\beta}$ as a $2\times2$
matrix, where $\alpha,\beta$ can take on the values $l$ and $t$, for
the longitudinal and transverse directions with respect to ${\bf q}$.
The density-density response function $K_{\rho\rho}^e$ is related to
$K^e_{\alpha\beta}$ by current conservation and gauge invariance, so
that $K^e_{\rho\rho}({\bf q},\omega)= (q^2/\omega^2) K^e_{ll}({\bf
  q},\omega)$.  

We simplify the action by the approximation $\bar\newpsi\newpsi\to n_0$ in
the diamagnetic terms with $n_0$ the average electron density. This
simplification is implicit in the random-phase approximation used by
HLR and was also used by MS.  It becomes exact in the limit where the
momentum cutoff Q, discussed in the Introduction and in Sec. 
\ref{sec:smallQ} below, is
taken to be very small.

As a first step towards an effective theory of the low-energy
quasiparticles, we integrate out the CS field ${\bf a}$, which
describes the magnetoplasma modes, and obtain 
\begin{eqnarray}
 \label{eff-action}
   S=\bar\newpsi&& i\partial_0\newpsi-{1\over2m}|\nabla\newpsi|^2
    -{1\over2m^2}{\bf g}\,{U\over{1-n_0U/m}}\,{\bf g}
    \nonumber\\
    &&-{1\over m}{\bf A}\,{1\over{1-n_0U/m}}\,{\bf g}-{n_0\over2m}{\bf A}\,
    {1\over{1-n_0U/m}}\,{\bf A}.
\end{eqnarray}
Here, space and time integrals are left implicit, ${\bf g}$ denotes
the canonical momentum density of the fermions, 
${\bf g} = -(i/2) (\bar\newpsi \nabla \newpsi - (\nabla \bar\newpsi) \newpsi )$,
and the operator $U$,
in frequency and momentum representation, is
\begin{equation}
   U({\bf q},\omega)=\left(\begin{array}{cc}(q^2/\omega^2)v(q) & 
   2\pi\tilde\phi/i\omega \\
   -2\pi\tilde\phi/i\omega &0\end{array}\right)_.
\end{equation}

Differentiating the effective action (\ref{eff-action}) with respect
to the source field ${\bf A}$, we obtain for the electronic current
response function a matrix $K^e$, which we write in the form
$K^e=K^{\rm mp}+K^d$ with
\begin{equation}
 \label{mpk}
    K^{\rm mp}({\bf q},\omega)=-{n_0\over m}\left[1-{n_0\over m}U
      \right]^{-1}
\end{equation}
and 
\begin{eqnarray}
 \label{dk}
  & & K^d_{\alpha\beta}({\bf q},\omega) =   \\ & & \nonumber
\left\langle
    \left[{-1/m \over1-n_0U/m}\,{\bf g}({\bf q},\omega)\right]_\alpha
   \left[{-1/m \over1-n_0U/m}\,{\bf g}({\bf -q},-\omega)\right]_\beta\right\rangle_.
\end{eqnarray}
%
%
%
%
Since we are interested in response functions, here and below expressions like 
$\langle A_{{\bf q},\omega}B_{-{\bf
    q},-\omega}\rangle$ are to be understood as  the Fourier transform of the 
retarded correlator  $i\theta (t)\langle [A({\bf x},t), B(0,0)] \rangle$. 
In this section, angular brackets denote an expectation value in the ground 
state of the action (\ref{eff-action}) with ${\bf A} = 0$. 
The superscripts mp and $d$ anticipate the fact that the two
contributions to $K^e$ will be identified with the magnetoplasma
oscillators and the dipolar quasiparticles, respectively.

We first consider $K^{\rm mp}$ in more detail.  One readily
establishes that the corresponding contribution to the electronic
density-density correlator is
\begin{equation}
   K^e_{\rho\rho}({\bf q},\omega)={n_0\over m}
      {q^2\over \omega_c^2-\omega^2+{n_0\over m}q^2 v(q)}.
\end{equation}
Hence, this contribution to $K^e$ reproduces Kohn's mode.  The
corresponding result for the conductivity tensor for small ${\bf q}$
and $\omega$ reproduces the correct Hall conductivity,
\begin{equation}
   \sigma^{\rm mp}({\bf q},\omega)={1\over2\pi\tilde\phi}
   \left(\begin{array}{cc}i\omega/\omega_c& 
   1\\
   -1&i\omega/\omega_c\end{array}\right)_.
\end{equation}
These results reflect the facts that there is no contribution to the
Hall conductivity from $K^d$ at $\nu = 1/2$ in a system without
impurities, and that the contribution of the quasiparticles to the
electromagnetic response is negligible compared to that of the Kohn
mode, for any finite frequency, in the limit $q \to 0$.

Nevertheless, the fermions are responsible for interesting low-energy
physics. Before deriving an effective low-energy action, it is useful
to collect the conditions which such an approximation would need to
satisfy:

A. The first condition (condition A) arises because, due to gauge
invariance, we can compute the fermionic contribution to the
density-density correlator $K^e_{\rho\rho}$ in two different ways. We
can either use directly that $\rho^e=\rho=\bar\newpsi\newpsi$, or compute
$K^e_{ll}$ from Eq.\ (\ref{dk}) and then use its relation to
$K^e_{\rho\rho}$. We will see below that assuring that both approaches
yield the same result is closely related to a consistent treatment of
the constraint in the Hamiltonian approach.

B. The second condition (condition B) is a
consequence of the fact that even once the Weyl gauge $a_0=0$ is
specified, we can still make the limited gauge transformations
\begin{eqnarray}
     a_l({\bf q},\omega=0)&\to&a_l({\bf q},\omega=0)+f({\bf q})
    \nonumber\\
   a_t({\bf q}=0,\omega=0)&\to&a_t({\bf q}=0,\omega=0)+{\rm const}. 
\label{eq:fif}
\end{eqnarray}
These transformations must leave the action and the physical current 
unchanged provided that we simultaneously change the 
canonical momentum density of the fermions by
\begin{eqnarray}
   g_l({\bf q},\omega=0)&\to&g_l({\bf q},\omega=0)-n_0f({\bf q})
    \nonumber\\
   g_t({\bf q}=0,\omega=0)&\to&g_t({\bf q}=0,\omega=0)-n_0\,{\rm const}.
\label{shiftg}
\end{eqnarray}
Here, we made the replacement $\bar\newpsi\newpsi\to n_0$ as in the
diamagnetic term of the action.  For $q=0$ the transformation
(\ref{shiftg}) shifts the momentum of each fermion by a constant.  We
shall shortly employ an effective low-energy action where the the
field ${\bf a}$ no longer appears (See Eq. (\ref{low-e-action}) below
which results from Eq. (\ref{eff-action})). In this case, the
transformation (\ref{shiftg}) by itself, without (\ref{eq:fif}), must
leave the action and the physical currents unchanged. This is what we
have called ${\bf K}$-invariance. As a consequence of this invariance,
the zero frequency correlator of the momentum density must diverge as
$q\to0$, being the inverse of the energy cost associated with a
uniform momentum shift.

We now consider the action (\ref{eff-action}), for ${\bf A} = 0 $, in
the limit of small frequency and wavevector. In this limit, we can
expand in powers of $m U^{-1}/n_0$ whose matrix elements are proportional to
either $q$ or $\omega$. This gives the effective low-energy action
\begin{equation}
\label{low-e-action}
  S=\bar\newpsi i\partial_0\newpsi-{1\over2m}|\nabla\newpsi|^2
   +{1\over2n_0m}{\bf g}\,{\bf g}+{1\over 2n_0^2}{\bf g}\,U^{-1}\,{\bf g}.
\end{equation}
Applying the same expansion to $K^d$, one obtains
\begin{equation}
\label{dipole-jj}
    K^d_{\alpha\beta}({\bf q},\omega)\simeq\left\langle\left[{1\over n_0}U^{-1}
   \,{\bf g}\right]_\alpha
   \left[{1\over n_0}U^{-1}\,{\bf g}\right]_\beta\right\rangle_.
\end{equation}
This implies that a fermionic momentum density ${\bf g}$ is associated
with a charge current
\begin{equation}
\label{dipole-current}
   {\bf j}^e={1\over n_0}U^{-1}{\bf g}=\left[\begin{array}{c}
       -{i\omega\over2\pi\tilde\phi n_0}\,g_t\\
       {i\omega\over2\pi\tilde\phi n_0}\,g_l-
       {v(q)q^2\over(2\pi\tilde\phi)^2n_0}\,g_t
       \end{array}\right]_.
\end{equation}
The (electronic) continuity equation yields for the charge density
\begin{equation}
\label{dipole-density}
   \rho^e({\bf q},\omega)={iq\over2\pi\tilde\phi n_0}
     g_t({\bf q},\omega).
\end{equation}
These equations are naturally interpreted if one identifies the
fermions as dipoles with dipole moment perpendicular to their canonical 
momentum. This is illustrated in Fig. 1. We
see that the time (space) derivative of a transverse momentum density 
is associated with a longitudinal charge current (charge density).
Similarly, one convinces oneself that the time derivative of a
longitudinal quasiparticle momentum corresponds to a transverse charge
current.  The additional contribution to the transverse charge current
associated with the Coulomb interaction can be understood as follows.
According to Eq.\ (\ref{dipole-density}), a transverse fermion momentum-density
is associated with a charge density, which in turn produces 
an electric field.  Due to the finite Hall conductivity, this
longitudinal electric field induces a transverse
Hall current.

The above discussion, which was restricted to the contributions of
fluctuations at long wavelengths, should be directly applicable to the
model where there is a wavevector cutoff $Q$ which is taken to zero, 
a model in which the RPA is presumably exact (see Sec.\ref{sec:smallQ}).
However, there is another way of looking at the expansion leading to
the low-energy action (\ref{low-e-action}), which suggests that it
should have wider applicability.  As was discussed in the
introduction, we expect that in the limit where the band mass $m \to
0$, so that the cyclotron frequency is infinite, there remains a
finite energy scale for intra-Landau-level excitations, and there
should be a finite contribution to the response functions from these
excitations. If one expands the effective action (\ref{eff-action}) in
powers of $m$, keeping only terms which do not vanish in the limit of
$m\to0$, one readily recovers from (\ref{eff-action}) the low-energy
action (\ref{low-e-action}).  While neither of these expansions is
truly satisfactory, it is encouraging that both lead to the same
result.

\begin{figure}
\label{fig-dipoles}
\iflongversion
  {\bf Figure 1 Goes here \\}
\else
\centerline{\epsfig{figure=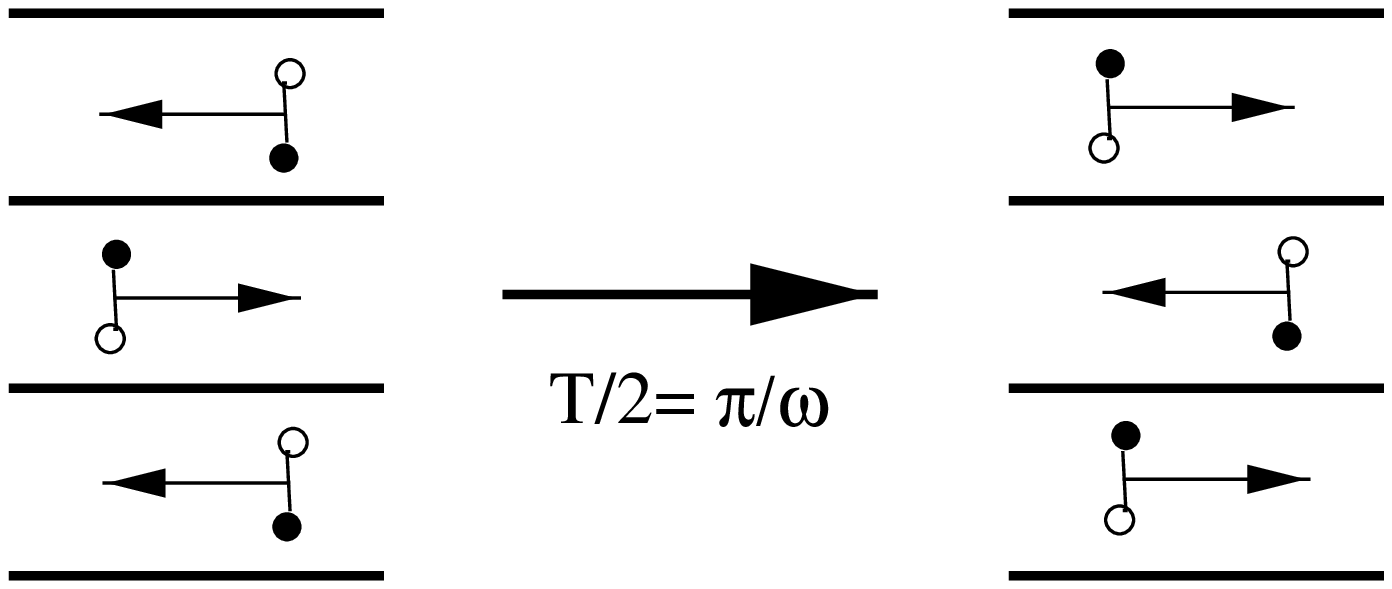,width=7.5cm,angle=270}}
\fi
\mycaption{{\bf Fig. 1}: Schematic drawing of a transverse
  quasiparticle momentum density $g_{t}=g_0{\protect\cos}
  (qy-{\protect\omega} t)$ at two times separated by half a period.
  Full (empty) circles indicate positive (negative) charges and arrows
  the direction of the quasiparticle momentum.  In agreement with
  Eqs.\protect\ 
  (\protect\ref{dipole-current},\protect\ref{dipole-density}), the
  gradient of a transverse momentum density is associated with a
  charge density and its time derivative implies a longitudinal charge
  current.}
\end{figure}

An important feature of the quasiparticle action (\ref{low-e-action})  is
that it contains a coupling between longitudinal and transverse momentum
densities due to the off-diagonal terms in ${\bf g}U^{-1}{\bf g}$. This
additional term is crucial in order for the action Eq.\
(\ref{low-e-action}) to satisfy condition A.  To see this, one can derive
the analog of the continuity equation for the dipole action
(\ref{low-e-action}). The continuity equation is modified in the present
case because the interactions involve the momentum currents ${\bf g}$.
Within the saddle-point approximation, we can proceed
either by directly applying the Euler--Lagrange equations of
motion or by deriving the Noether current associated with the invariance
of the action under a global phase change of $\psi$. One thus obtains
\begin{equation}
  i\omega\left\{\rho({\bf q},\omega)-{iq\over2\pi\tilde\phi n_0}
     g_t({\bf q},\omega)\right\}=0.  \end{equation}
In deriving this
equation, we have again made the replacement $\bar\psi\psi\to n_0$. Hence,
we observe that for non-zero frequency and at the RPA level, this modified
continuity equation guarantees that condition A is satisfied.  By
contrast, if one
neglects the coupling of longitudinal and transverse momentum densities,
one finds $i\omega\rho({\bf q},\omega)=0$, in violation of condition A.
Indeed, we will see below that this coupling is crucial for precisely
reproducing the RPA results of HLR from the present approach.

We now turn to obtaining the dipole contribution to the electronic
density-density correlator by using Eq.\ (\ref{dipole-jj}). To this
end, we compute the quasiparticle momentum-density correlation function
$D^d_{\alpha\beta}({\bf q},\omega)=\langle g_\alpha({\bf q},\omega)
g_\beta(-{\bf q},-\omega)\rangle$ in the random-phase approximation. One
finds
\begin{equation}
\label{dipole-rpa}
   [(D^d)^{-1}]_{\alpha\beta}=[(D^0)^{-1}]_{\alpha\beta}
     -{1\over n_0m}\delta_{\alpha\beta}
      -{1\over n_0^2}[U^{-1}]_{\alpha\beta},
\end{equation}
where $D^0$ denotes the correlator of free fermions.  In the limit
$\omega\ll v_Fq$, we have
\begin{eqnarray}
\label{free-fermion-jj}\nonumber
    {1\over m^2}D^0_{ll}(q,\omega)&=&{n_0\over m}
     +{\omega^2\over q^2}\left[{m\over 2\pi}
     +{m\over2\pi}{i\omega\over v_Fq}\right]\\
     {1\over m^2}D^0_{tt}(q,\omega)&=&{n_0\over m}-{q^2\over 24\pi m}
      +{i2n_0\omega\over \kf q}.
\end{eqnarray}
To bring out the unusual properties of the dipoles, we first focus on
the limit of zero frequency. In this limit, $D^0_{ll}({\bf
  q},\omega=0)=n_0m$ and $D^0_{tt}({\bf q},\omega=0)=n_0m
-{mq^2\over 24\pi}$. The result for $D^0_{ll}({\bf q},\omega=0)$ 
follows directly
from gauge invariance, specifically the fact that a {\it
  time-independent} longitudinal vector potential can have no physical
effect, and the usual expression for the physical current operator.

Using Eqs.\ (\ref{dipole-rpa},\ref{free-fermion-jj}), one finds 
for $\omega=0$ that $D^d$ of the dipoles is diagonal with
\begin{eqnarray}
   [(D^d)^{-1}({\bf q},\omega=0)]_{ll}&=& 0\nonumber \\
   {}[(D^d)^{-1}({\bf q},\omega=0)]_{tt}&=&{q^2\over 24\pi mn_0^2}
     +{q^2v(q)\over(2\pi\tilde\phi)^2n^2}.
\end{eqnarray}
Hence, the longitudinal correlator of the dipoles diverges for all
${\bf q}$ once $\omega=0$, while the transverse correlator diverges
for ${\bf q}\to 0$ like $1/q$ for Coulomb interaction and like $1/q^2$
for a short-range interaction, in stark contrast to the free-fermion
results quoted above. These results are a consequence of
condition B. We conclude that the dipole action (\ref{low-e-action})
satisfies condition B at least at RPA level.

How do the dipoles evade the usual theorem that the zero-frequency
correlator $D_{ll}({\bf q},\omega=0)$ be $n_0m$?  To make the dipole
action (\ref{low-e-action}) gauge invariant, we would need to replace
the canonical momenta in the interaction terms by minimally coupled
kinetic momenta. This introduces the vector potential into the
interaction terms. Computing the gauge-invariant current operator by
taking the derivative of this Hamiltonian with respect to the vector
potential, we see that it takes a form different from the usual one.
It is for this reason that the standard argument for $D^0_{ll}({\bf
  q},\omega=0)$ does not apply to the dipoles.

Using Eq. (\ref{dipole-density}) and the RPA Eqs.  (\ref{dipole-rpa}
,\ref{free-fermion-jj}), one finds for the electronic density-density
correlator in the limit $\omega\ll v_Fq$
\begin{equation}
\label{kee}
   K^e_{\rho\rho}({\bf q},\omega)
      ={1\over v(q)+{2\pi\over m}+{(2\pi\tilde\phi)^2\over 24\pi m}
      -i(2\pi\tilde\phi)^2{2n_0\omega\over \kf q^3}}.
\end{equation}
This coincides exactly with the corresponding expression 
obtained by HLR.\cite{HLR} One easily checks that the second term in
the denominator is due to the coupling between longitudinal and
transverse momentum density in the above dipole action.

It is instructive to compare these calculations to those in the
Coulomb gauge $\nabla\cdot{\bf a}=0$ used by HLR. In that gauge, the action
is
\begin{eqnarray}
\label{Coulomb-action}
   S=&&\int dt\,d{\bf r}\left\{{\bar\newpsi}i\partial_0\newpsi
    +a_0\left(-{\nabla\times{\bf a}\over 2\pi
    \tilde\phi}+{\bar\newpsi}\newpsi-n_0\right)\right.
    \nonumber\\
    &&\left.-{1\over 2m}{\bar\newpsi}
    (-i\nabla+{\bf A}+{\bf a})^2\newpsi\right\}
    +S_{\rm Coul}.
\end{eqnarray}
While the CS field ${\bf a}$ has only a transverse component, we keep
both longitudinal and transverse component for the source field ${\bf
  A}$. Integrating out the CS field, one obtains the effective
fermionic action
\begin{eqnarray}
    S=\bar\newpsi&&\partial_0\newpsi-{1\over 2m}|\nabla\newpsi|^2-{1\over
      m}{\bf Ag}-{n_0\over2m}{\bf A}^2
       \nonumber\\
      &&-{1\over2}(\bar\newpsi\newpsi-n_0)\left(v(q)+{(2\pi\tilde\phi)^2n_0\over
        mq^2}\right)(\bar\newpsi\newpsi-n_0)
       \nonumber\\
      &&-(\bar\newpsi\newpsi-n_0){2\pi\tilde\phi\over iq}\left({1\over
        m}g_t+{n_0\over m}A_t\right)
\end{eqnarray}
This effective action does not suggest a simple low-energy expansion
analogous to the one found above for the Weyl gauge.  Differentiating
$S$
with respect to the source field ${\bf A}$, we obtain for charge
density and current
\begin{eqnarray}
   \rho^e&=&\bar\newpsi\newpsi \nonumber \\
   j_l^e&=&-{1\over m}g_l  \nonumber\\
   j_t^e&=&-{1\over m}g_t-{n_0\over m}a_t
\end{eqnarray}
where $a_t=(2\pi\tilde\phi/iq)(\bar\newpsi\newpsi-n_0)$. These expressions
are appropriate for composite fermions which carry charge $e$ and
which are subject to an effective magnetic field associated with
deviations in density from half filling. The vector potential $a_t$ appears 
explicitly in the formula for the electron current, and there is no manifest 
separation between the high energy and low energy physics at this stage.

To summarize, by analyzing the low energy response of the Chern-Simons
Lagrangian in the Weyl gauge within the RPA we identified
quasiparticles of a dipolar nature, whose position coincides with that
of the electrons. In the next section we carry out an analysis of the
Chern-Simons Hamiltonian that leads us to describe the same low energy
dynamics in terms of quasiparticles whose position is shifted with
respect to that of the electrons.

\section{Shifted quasi-particles: a Hamiltonian approach}
\label{sec:Hamilton}

The Hamiltonian (\ref{ham}) represents the problem in terms of two
coupled sets of degrees of freedom, $\newpsi_\cs$ and $\bf a_\cs$.  The
dipolar field-theoretic approach to this Hamiltonian, initiated by
Murthy and Shankar\cite{Shankar}, is motivated by the hope that if
transformed properly, the gauge field degrees of freedom should be
decoupled from the fermionic degrees of freedom at the long wavelength
limit.  The former would then describe a set of harmonic oscillators
representing the inter Landau level magneto plasmons, while the latter
would describe intra Landau level physics.  An insight into this long
wavelength decoupling, as well as into the $K$--invariance, is gained
when the Hamiltonian (\ref{ham}) is written as
\begin{eqnarray} \label{ham2}
\nonumber
H_\cs&=&\int d\vec r\left[ \frac{1}{2m}|\nabla\newpsi_{\cs}|^2
 - {\bf g}_\cs \left( \frac{1}{2mn_\cs} \right) {\bf g}_\cs
\right.  \\ \nonumber
 &+&
 \left. 
\Big( {\bf a}_\cs-
{\bf g_{\cs}}\frac{1}{n_\cs}\Big )\frac{n_\cs}{2m} \Big( {\bf a}_\cs-
\frac{1}{n_\cs}{\bf g_{\cs}}\Big )\right]\\ 
+\frac{1}{2}
\int &d\vec r& \int d\vec{r'}(n_\cs({\bf r})-n_0)v({\bf r}-{\bf r'})
(n_{\cs}({\bf r'})-n_0) 
\label{hamrr}
\end{eqnarray}
In Eq. (\ref{hamrr}), the momentum density of the fermions is $ {\vec
  g}_\cs({\bf r})\equiv -(i/2)
[\newpsidagger_\cs \nabla\newpsi_\cs-(\nabla\newpsidagger_\cs)
\newpsi_\cs] $, their density is $n_{\cs}({\bf
  r})\equiv\newpsidagger_\cs({\bf r})\newpsi_\cs({\bf r})$, the $\bf
r$ dependence of $\newpsi,{\bf g}$ and $n$ is suppressed for clarity.

Given the form of the Hamiltonian, one is led to approximate
$n_{\cs}\approx n_0$ inside the square brackets of (\ref{hamrr}), and
to write the Hamiltonian in momentum space as (note that the area of
the system is taken to be $1$)
\begin{eqnarray} \nonumber
&H_\cs& \approx  \sum_{\bf k} \frac{k^2}{2m}\newpsi_\cs ({\bf 
k})\newpsi_{\cs}({\bf -k}) \\ \nonumber
 &-& \frac{1}{2mn_0}\sum_{\bf q}{\bf g}_\cs({\bf q}){\bf
   g}_\cs({\bf
   -q})+\sum_{\bf q\ne 0}n_\cs({\bf q})v(q)n_\cs(-{\bf q})\\ 
 &+& \sum_{\bf q}\frac{n_0}{2m}\Big | {\bf a}_\cs({\bf q})-
\frac{1}{n_0}{\bf g}_\cs({\bf q})\Big |^2 
 \label{hamq}
\end{eqnarray}
Our small $Q$ model is the Hamiltonian (\ref{hamq}) with all the sums
over $q$ ranging between $0<q<Q$, and $Q$ treated as a small
parameter.  Physically, this model amounts to considering the charge
and the flux tubes of the composite fermions as smeared over distance
$\sim Q^{-1}$.  Our analysis below focuses on the kinetic part of the
Hamiltonian. Thus, at this stage we omit the electron-electron
interaction part ($v(q)$). Its inclusion in the analysis below is
straightforward.

Within the small $Q$ model, we employ the unitary
transformation suggested by Murthy and Shankar\cite{Shankar},
\begin{equation}
U=\exp{i\frac{1}{2\pi{\tilde\phi}n_0}\sum_{\bf q}^{\scriptstyle Q} 
\ {\bf g}_{\bf q}\times {\bf a}_{-{\bf
    q}}}
\label{U}
\end{equation}
in order to redefine the gauge field coordinates to be ${\bf
a}_\cs-\frac{1}{n_0}{\bf g_\cs}$, and make the coupling between the
redefined gauge field and fermion coordinates vanish in the long
wavelength limit.  Note that in Eq. (\ref{U}), as in (\ref{hamq}), the
sum over $q$ is limited to $0\le q<Q$. In our notation the
operators before the transformation have a subscript CS, while this
subscript is missing from their transformed counterparts. For example,
the transformed Hamiltonian $H$ is defined by $H_\cs=U^\dagger HU$.
 
Let us first consider the case $Q=0$ in which the sums in Eqs.
(\ref{hamq}) and (\ref{U}) are limited to the $q=0$ term.  The
fermions and the oscillators are exactly decoupled by the
transformation, and the transformed Hamiltonian can be written as,
\begin{equation}
\label{eq:haldane}
  H_{Q=0}=
  \frac{1}{2m} \left|(-i{\bf \nabla} - {\bf g}_{q=0})\newpsi\right|^2+
\frac{n_0}{2m}{\bf a_{q=0}}^2.
\end{equation}
The Hamiltonian (\ref{eq:haldane}) is manifestly $K$--invariant. The
source of that invariance is the invariance of the untransformed
Hamiltonian (\ref{ham2}) to the gauge transformation
\begin{eqnarray}
{\bf a}_\cs({\bf r})&\rightarrow& {\bf a}_\cs({\bf r}) +{\bf K}\nonumber
\\
\newpsi_\cs({\bf r})&\rightarrow& e^{i{\bf K}\cdot {\bf r}}\newpsi_\cs({\bf
  r}). 
\label{gauge}
\end{eqnarray}
This gauge transformation, which is analogous to Eqs. (\ref{eq:fif}) and
(\ref{shiftg}), leaves the redefined gauge field coordinate
${\bf a}={\bf a}_\cs-\frac{1}{n_0}{\bf g_\cs}$ unchanged.  Thus it
leaves unchanged the energy stored in magnetoplasmons. Therefore, for
the total energy to be gauge invariant {\it the energy stored in the
  transformed fermionic degrees of freedom should be also invariant
  under the gauge transformation, which shifts the momentum of each
  fermion by $\bf K$.}  As we see below, for non zero $Q$ the unitary
transformation that redefines ${\bf a}_\cs$ redefines the fermions and
introduces a coupling between the transformed fermions and gauge
fields.  However, it is still true that in the small $Q$ limit the
invariance of the pre-transformed Hamiltonian $H_\cs$ to the
transformation (\ref{gauge}) implies the $\bf K$-invariance of the
transformed fermions.

A Hamiltonian similar to the first term in  
(\ref{eq:haldane}), with the bare mass
replaced by a renormalized one, was suggested by Haldane as an
effective Hamiltonian for low energy excitations of the $\nu=1/2$
state \cite{Haldane1}.

For a small but non-zero $Q$ the momentum current ${\bf g}_{\bf q}$
does not commute with other fermionic operators appearing in the
Hamiltonian. Consequently we are unable to calculate the transformed
Hamiltonian exactly, but use the following small $Q$ approximation: we
neglect all terms in the transformed Hamiltonian that involve more
than one integral over $q$, and calculate all response functions by
neglecting all diagrams that contain interaction lines in which the
momentum exchange is integrated.  These diagrams are neglected since
the integration over the momentum exchange is of the form $\int^Q
d{\bf q}$ and yields terms of higher order in $Q$. Their neglect
constitutes the random phase approximation (RPA).

As is well known, the RPA can be defined either by a diagrammatic
classification, as given above, or, equivalently, by approximating
commutation relations. The operators whose commutation relation are
relevant for applying the unitary transformation at hand are the
density $\rho_{\bf q}$, the Cartesian components of the momentum
density $g_{\alpha,{\bf q}}$ and the Cartesian components of the
vector ${\bf C}_{\bf q}$, which is defined by
\begin{equation}
{\bf C}_{\bf q}\equiv i[H_K,{\bf g}_{\bf q}]=
i\sum_{\bf k}\frac{{\bf q}\cdot {\bf k}}{m} 
{\bf k}\newpsidagger_{{\bf k}+\frac{\bf 
q}{2}}\newpsi_{{\bf k}-\frac{\bf q}{2}}
\label{cofq}
\end{equation}
with $H_K\equiv\sum_{\bf k}\frac{k^2}{2m}\newpsidagger_{\bf k}\newpsi_{\bf 
k}$. 

The approximate commutation relations between these operators are:
\begin{eqnarray}\label{appro}
[\rho_{\bf q},{\bf g}_{\bf q'}]&=&{\bf q}\rho_{{\bf q}+{\bf q'}}
\approx {\bf q}n_0\delta_{{\bf q},-{\bf q'}} 
\\
{[}g_{\alpha,{\bf q}},g_{\beta,{\bf q'}}]&=&
q'_\alpha g_{{\mathscriptsize \beta},{\bf q}+{\bf
  q'}}-q_{\mathscriptsize \beta} g_{\alpha,{\bf q}+{\bf q'}}\approx 0
\\
{[}g_{l,{\bf q}},C_{l,{\bf q'}}] &\equiv&{[}{\hat q}\cdot{\vec g_{\bf q}},{\hat 
q'}\cdot{\vec C_{\bf 
q'}}]\approx
i\frac{3\pi n_0^2q^2}{m}\delta_{{\bf q},-{\bf q'}} 
\\
{[}g_{t,{\bf q}},C_{t,{\bf q'}}]&\equiv&{[}{\hat q}\times{{\bf g}_{\bf q}},{\hat 
q'}\times{\vec C_{\bf 
q'}}]\approx
i\frac{\pi n_0^2q^2}{m}\delta_{{\bf q},-{\bf q'}}\\
{[}g_{l,{\bf q}},C_{t,{\bf q'}}]&=&[g_{t,{\bf q}},C_{l,{\bf q'}}]=
[C_{\alpha,{\bf q}},C_{{\mathscriptsize \beta},{\bf q'}}]=0
\end{eqnarray}

Within this approximation scheme, the following simple relations hold
between the pre- and post- transformation operators:
\begin{eqnarray}
{\bf a}_\cs(\bf q)&=&{\bf a}_{\bf q}+\frac{1}{n_0}{\bf
   g}_{\bf q}\nonumber \\
\rho_\cs({\bf q})&=&\rho_{\bf q}+i\frac{{\bf q}\times {\bf a}_{\bf
  q}}{2\pi\tilde\phi}+\frac{i}{2}
\frac{{\bf
    q}\times{\bf g}_{\bf q}}{2\pi{\tilde \phi}n_0}\nonumber \\
{\bf g}_\cs({\bf q})&=&{\bf g}_{\bf q}
\label{prepost}
\end{eqnarray}
The transformation of $[H_K]_\cs$ is more complicated.  Carrying out
the transformation to lowest order in $Q$ requires an expansion of the
exponent (\ref{U}) to fourth order. The resulting transformation is
the following, with each line giving the contribution of one order in
the expansion,
\iflongversion
\typeout{Skippingmulticols}
\else
\end{multicols}
\fi
\begin{eqnarray}
{[}H_K]_\cs 
&=& H_K\\
&+&
\frac{1}{(2\pi\tilde\phi)n_0}
\sum_{\bf q}^{\myscriptsize Q}
{\bf C}_{\bf q}\times {\bf  a}_{-\bf q}\\
&-&
\frac{1}{2(2\pi\tilde\phi)n_0^2}\sum_{\bf q}^{\myscriptsize Q}
 {\bf C}_{\bf q}\times
{\bf
  g}_{-\bf q}+
\frac{\pi }{2(2\pi\tilde\phi)^2 m}\sum_{\bf q}^{\myscriptsize Q}
[3|{\bf q}\times {\bf a}_{\bf q}|^2+|{\bf q}\cdot {\bf a}_{\bf q}|^2]\\
&-&\frac{\pi}{2(2\pi{\tilde\phi})^2n_0m}\sum_{\bf q}^{\myscriptsize Q}
[3({\bf q}\times {\bf a_{\bf q}})\cdot
({\bf q}\times {\bf g}_{\bf -q}) +({\bf q}\cdot {\bf a}_{\bf q})
({\bf q}\cdot {\bf g}_{\bf -q})]
\\
&+&\frac{\pi}{8(2\pi{\tilde\phi})^2 n_0^2 m}\sum_{\bf q}^{\myscriptsize Q}
[3|{\bf q}\times{\bf g}_{\bf q}|^2+
|{\bf q}\cdot{\bf g}_{\bf q}|^2]
\label{prepostb}
\end{eqnarray}
We now use Eqs. (\ref{prepost}) -- (\ref{prepostb}) 
to calculate the transformed form of
the Hamiltonian and the constraint. The Hamiltonian is,  
\begin{eqnarray} \nonumber
H({\rm small\ }Q)&\approx&\sum_{\bf k}
\frac{k^2}{2m}\newpsidagger_{\bf k}\newpsi_{\bf
  k}
 - \frac{1}{2mn_0}\sum_{\bf q}^{\myscriptsize Q}
{\bf g}_{\bf q}{\bf g}_{\bf
   -q}
  + \frac{\pi}{8(2\pi{\tilde\phi})^2 n_0^2 m}\sum_{\bf q}^{\myscriptsize Q}
  [3|{\bf q}\times{\bf g}_{\bf q}|^2+
|{\bf q}\cdot{\bf g}_{\bf q}|^2]\\ &  
  -&\frac{1}{2(2\pi\tilde\phi)n_0^2}\sum_{\bf q}^{\myscriptsize Q}
 {\bf C}_{\bf q}\times
{\bf
  g}_{-\bf q}  \nonumber\\
&+& \frac{1}{(2\pi\tilde\phi)n_0}\sum_{\bf q}^{\myscriptsize Q}
{\bf C}_{\bf q}\times {\bf
  a}_{-\bf q}
 -\frac{\pi}{2(2\pi{\tilde\phi})^2n_0m}[3({\bf q}\times {\bf a}_{\bf q})\cdot
({\bf q}\times {\bf g}_{\bf -q}) +({\bf q}\cdot {\bf a}_{\bf q})
({\bf q}\cdot {\bf g}_{\bf -q})] \nonumber\\
&+& \sum_{\bf q}^{\myscriptsize Q}
\frac{n_0}{2m}\Big | ({\bf a}_{\bf q}-{\bf A}_{\bf q})\Big |^2 
+\frac{\pi }{2(2\pi\tilde\phi)^2 m}\sum_{\bf q}^{\myscriptsize Q}
[3|{\bf q}\times {\bf a}_{\bf q}|^2+|{\bf q}\cdot {\bf a}_{\bf q}|^2]\nonumber 
\\
&+&\sum_{\bf q}^{\myscriptsize Q}
\Big [\rho({\bf q})+\frac{i{\bf q}\times {\bf a}({\bf
  q})}{2 \pi \tilde \phi} +\frac{i}{2^{\phantom{\dagger}}}
\frac{{\bf
    q}\times{\bf g}}{2\pi{\tilde \phi}n_0}\Big ]\ V_{\bf -q}
\label{hd}
\end{eqnarray}
\iflongversion
\typeout{Skippingmulticols}
\else
\begin{multicols}{2}
\fi
This Hamiltonian is composed of several terms.  Purely fermionic
terms include a kinetic energy, a current-current interaction energy
and an interaction term of the form ${\bf C}\times{\bf g}$.  The
current-current interaction has two terms, the second and third in
(\ref{hd}). The latter is small, by a factor of $q^2/n$, compared to
the former. Pure gauge field terms include the self energy of the
magneto-plasmons, which, too, has an additional term of order $q^2$.
There are two types of fermion-gauge field coupling terms, one
coupling ${\bf C}$ to ${\bf a}$, and one, of order $q^2$, coupling
${\bf g}$ to $\bf a$. Finally, we have incorporated in this
expression also a probing potential $(V_{\bf q},{\bf A}_{\bf q})$
(which may also be time dependent) for the future use of calculating
response functions.

The Chern-Simons constraint is transformed to the constraint 
(\ref{constraint}):
\begin{equation}
\rho({\bf q})=\frac{i}{2}
\frac{{\bf
    q}\times{\bf g}_{\bf q}}{2\pi{\tilde \phi}n_0}
\nonumber
\end{equation}
The transformed constraint commutes with the transformed Hamiltonian.
This observation reflects the consistency of our approximation scheme,
since the pre-transformed constraint commutes with the pre-transformed
Hamiltonian.

Eq. (\ref{constraint}) together with Eq. (\ref{prepost}) leads to the
identification of the electronic density as,
\begin{equation}
\rho_\cs({\bf q})=i\frac{{\bf
    q}\times{\bf g}_{\bf q}}{2\pi{\tilde \phi}n_0}+i\frac{{\bf q}\times {\bf 
a}_{\bf
  q}}{2\pi {\tilde\phi}}
\label{udens}
\end{equation}
The first term indicates that $\hat z \times {\bf g}/
(2\pi{\tilde\phi}n_0)$ is a dipolar field.  Thus, a transformed
fermion carrying a momentum ${\bf k}$ carries an electronic dipole
moment $e {\hat z}\times {\bf k}/(2 \pi \tilde \phi n_0)$, as one
expects from a dipole in a magnetic field \cite{KallinHalperin}. 
Since the spectrum of the oscillators is gapped, with the lowest
frequency being the electronic cyclotron frequency, their response to
a driving force of low frequency $\omega$ is small by a factor
$\omega/\omega_c$, compared to that of the dipoles.  Thus, the
contribution of the oscillators to the electronic charge density
(\ref{udens}) is negligible, and we may approximate $\rho_\cs({\bf
  q})\approx 2\rho({\bf q})= i {\bf q}\times{\bf g} / (2\pi{\tilde
  \phi}n_0)$. As explained in the introduction, the factor of $2$ between the 
electronic density $\rho_{CS}$ and the quasi-particle density $\rho$ indicates 
the shifting of the quasi-particle position from the electronic one.

The electronic physical current ${\bf j}^e$ 
may be identified by taking the derivative of the Hamiltonian (\ref{hd}) with 
respect to $\bf A$.
To leading order in $Q$, it is, 
\begin{equation}
{\bf j}^e=\frac{n_0}{m}( {\bf a}-{\bf A})
\label{current}
\end{equation}
It is expressed
in terms of the oscillators alone, and has a prefactor $n_0/m$. An
oscillators' response of order $\omega/\omega_c$ results, then, in an
electronic current that is independent of the mass.

We now turn to calculate response functions of the $\nu=1/2$ state
using the decoupled Hamiltonian (\ref{hd}), the Chern--Simons
constraint (\ref{constraint}) and the expression (\ref{udens}) for the
electronic density. Most particularly, we are interested in the
density response $\rho({\bf q},\omega)$ to a scalar potential $V({\bf
  q},\omega)$, so we may set ${\bf A}=0$. Furthermore, we are
interested in the limit of low frequency $\omega\ll\omega_c$ and long
wavelength $q\ll\kf$, in which the fermions' response is much stronger
than the oscillators'. Thus, we may set the magnetoplasmons frozen in
their ground states and replace ${\bf a}$ in the Hamiltonian
(\ref{hd}) by its expectation value, zero. We are then left with the
dipoles, whose density response to $V$ we calculate within RPA.
Generally, within RPA one first calculates the response functions of
free fermions, denoted by $\Pi$, and then approximates the response
function of the interacting fermions, $\cal K$, by,
\begin{equation}
{\cal K}^{-1}=\Pi^{-1}+{\cal V}
\label{rpa}
\end{equation}
where $\cal V$ is the fermion-fermion interaction.

In the present case $\Pi, {\cal V, K}$ are all $5\times 5$ matrices.
To understand that unusual dimension, we note that the interaction
terms in the Hamiltonian (\ref{hd}) couple ${\bf g}$ and ${\bf C}$,
and we are interested in the density response. Thus, $\Pi, {\cal V}$
and ${\cal K}$ must have $\rho, C_l, g_l, C_t, g_t$ entries,
which we label by indices $i,j$ running from 0 to 4..  The matrix
$\Pi$ is the response matrix of free fermions, namely fermions subject
to the Hamiltonian $H_K$. This generates a simple relation between
terms involving ${\bf g}$ and ${\bf C}$ in $\Pi$. The calculation of
the matrix elements of $\Pi$ is straightforward, and is given in the 
Appendix.
It results in,
\iflongversion
\typeout{Skippingmulticols}
\else 
\end{multicols}
\fi
\begin{equation}
\Pi=\left ( \begin{array}{ccccc}
\Pi_{00} &     iqn_0+i\frac{m\omega^2}{q}\Pi_{00}  &  
\frac{m\omega}{q}\Pi_{00} & 0 & 0 
\\
-iqn_0-i\frac{m\omega^2}{q}\Pi_{00} & \frac{3\pi (qn_0)^2}{m}+mn_0\omega^2 & 
-i\omega(mn_0+\frac{(m\omega)^2}{q^2}\Pi_{00})  & 0 & 0    \\
\frac{m\omega}{q}\Pi_{00} & i\omega(mn_0+\frac{(m\omega)^2}{q^2}\Pi_{00}) & 

mn_0+\frac{(m\omega)^2}{q^2}\Pi_{00} & 0 & 0 \\
0 & 0 & 0 & \frac{\pi (qn_0)^2}{m}+\omega^2\Pi_{tt}& -i\omega \Pi_{tt} \\
0 & 0 & 0 & i\omega \Pi_{tt} & \Pi_{tt} 
\end{array}
\right )
\label{pifive}
\end{equation}
In (\ref{pifive}) we used the notation $\Pi_{00}$ for the
density-density response functions of free fermions and $\Pi_{tt}$ for
the transverse current-current response functions of free fermions. In
the limit of small $q,\omega$, with $\omega/q\rightarrow 0$,
\begin{equation}
\begin{array}{ccc}
\Pi_{00}&=&
\label{free-fermion-jja}
    {m\over 2\pi}(1+
    {i\omega\over v_Fq})\\
\Pi_{tt}&=&
    {n_0 m}-{q^2m\over 24\pi}
      +{{i2n_0\omega m^2}\over {\kf q}}.
\end{array}         
\end{equation}
as one sees in (\ref{free-fermion-jj}). 

The interaction matrix $\cal V$ is read off from the Hamiltonian (Eq.
\ref{hd}):
\begin{equation}
{\cal V}=\left (
\begin{array}{ccccc}
0 & 0 & 0 & 0 & 0 \\
0 & 0 & 0 & 0 & \frac{1}{4\pi{\tilde\phi}n_0^2} \\
0 & 0 & -\frac{1}{mn_0} + \frac{q^2\pi}{4m (2\pi\tilde\phi n_0)^2} & 
-\frac{1}{4\pi{\tilde\phi}n_0^2} & 0 \\
0 & 0 &- \frac{1}{4\pi{\tilde\phi}n_0^2} & 0 & 0 \\
0 & \frac{1}{4\pi{\tilde\phi}n_0^2} & 0 & 0 & -\frac{1}{mn_0}+ 
\frac{3q^2\pi}{4m (2\pi\tilde\phi n_0)^2}
\end{array}\right )
\label{vint}
\end{equation}
\iflongversion
\typeout{Skippingmulticols}
\else
\begin{multicols}{2}
\fi
Combining Eqs. (\ref{rpa}), (\ref{free-fermion-jja}), (\ref{pifive})
and (\ref{vint}) to calculate the matrix $\cal K$, we find that 
as long as $\omega\ne 0$ the matrix 
elements of $\cal K$ satisfy the constraint (\ref{constraint}). Since the
shifted quasi-particle density is just half of the electronic density,
the electronic density-density response function is four times that of
the shifted quasi-particles, i.e., it is $4{\cal K}_{00}$.  Calculating 
that element, we find it to equal Eq. (\ref{kee}) with $v(q)$ turned to
zero.  Including the electron-electron interaction in the matrix $\cal
V$ reproduces exactly Eq. (\ref{kee}), which is the electronic
density-density response function calculated either by using electron
centered quasi-particles or by using the FCS approach. Thus our
Hamiltonian approach, which describes quasi-particles that are shifted
away from the electrons, yields the same response functions as our
Lagrangian approach, which described electron centered
quasi-particles.

The calculation outlined in the previous paragraph is carried out for
a non-zero $\omega$, and its $\omega\rightarrow 0$ limit yields the
FCS result for the static limit. The $\omega=0$ case can be directly
calculated from Eqs. (\ref{rpa}), (\ref{free-fermion-jja}),
(\ref{pifive}) and (\ref{vint}), but some care is necessary, as there
are two subtle points. First, at $\omega=0$, for all $q$, the matrix
${\cal K}^{-1}$ has a zero eigenvalue. This eigenvalue corresponds to
a fluctuation in $g_l$, coupled to the $ C_t$. Thus, when inverting
the matrix, one must work in the subspace orthogonal to this
eigenmode, which is the $3\times 3$ subspace spanned by $\rho,g_t,
C_l$.  The infinite susceptibility corresponding to the zero
eigenvalue is not a problem, because there is no physical observable
or force field which couples to the mode.  The mode can be understood
as a remnant of the gauge invariance of the original problem, where an
arbitrary longitudinal momentum density could be added and compensated
by a change in the longitudinal ${\bf a}_\cs$, with no change in
energy or any other physical property.  A similar zero mode was found
in Section II, in the simpler case of a 2$\times$2 matrix.

The second subtle point is that the commutativity of the constraint
(\ref{constraint}) with the Hamiltonian (\ref{hd}) causes the
constraint to be satisfied automatically only for $\omega\ne 0$.
Exactly at $\omega=0$, the constraint needs to be imposed explicitly,
as by adding a term to the energy which becomes very large if the
constraint is violated.  Specifically, we add to the matrix ${\cal
  V}$, defined in (\ref{vint}), a matrix ${\cal V}^ \prime _{ij}=
\lambda U^*_i U_j$, with
\begin{equation}
U_j = \delta_{j0} - i q \delta_{j4}  / ( 4 \pi 
{\tilde \phi} n_0)
\label{Uj}
\end{equation}
and $\lambda \to \infty $.

The agreement we find between the density-density response functions
calculated using the dipolar quasiparticle approach and the FCS
approach also extends to other physical quantities, such as the
quasiparticle effective mass, as will be discussed in Section V
below.  First, however, we shall discuss further the physical meaning
and some mathematical consequences of the small $Q$ model, in Section
IV.

Before concluding this section, we note that unlike the $Q=0$
Hamiltonian, the Hamiltonian (\ref{hd}) is not exactly $K$-invariant.
A boost of the fermions' momentum by ${\bf K}$ does affect their
energy, but this effect is of high order in $Q$. To find out what that
order is, we note that shifting the momentum of each particle by $\bf
K$ amounts to the shift $\vec g(\vec q) \rightarrow \vec g(\vec q) +
\vec K \rho(\vec q)$.  With this shift, we find that the Hamiltonian
(\ref{hd}) acquires some new components, of which the most important
one is,
\begin{equation}
   - \sum_{\vec q \ne 0}^Q 
  \left[ \frac{\vec K \cdot \vec g_{\vec q} \rho_{-\vec q}}{
   mn_0} + \frac{K^2 \rho_{\vec q} \rho_{-\vec q}}{2 m n_0}\right]
\end{equation}
Due to isotropy, the expectation value of the first term must
vanish if the unshifted state is centered around zero momentum.  
As for the second term, its expectation value is
\begin{equation}
 \delta E(K) =  -\sum_{\vec q \ne 0}^Q \frac{K^2}{2 m} \langle
 \rho_{\vec q} \rho_{-\vec q} \rangle  = 
-\sum_{\vec q \ne 0}^Q \frac{K^2}{2 mn_0} S(q)
\end{equation}
For a Fermi liquid of particles interacting via short-range
interaction in two dimension, $S(q) \sim q$ so that $\delta E(K) \sim
Q^3$.  In the presence of interactions of longer range, the power of
$Q$ gets even larger.  The current--current interaction between the
fermions at hand makes this power larger as well.
Thus, the Hamiltonian (\ref{hd})
satisfies the $\vec K$-invariance,
at least to order $Q^2$, for small $Q$; but it does not do so at high
orders in $Q$, and therefore violates $\vec K$-invariance seriously for 
large values of the  cutoff $Q$.

\section{Comments on the small $Q$ limit}
\label{sec:smallQ}
In this section we make several comments regarding the small $Q$ model
we use.  First, it is instructive to identify the electronic problem
whose composite fermion formulation is our small $Q$ model. In the
small $Q$ composite fermion problem each fermion carries two flux
quanta anti-parallel to the external magnetic field, as well as a
charge $-1$, both smeared over a distance $Q^{-1}$. To transform back
from composite fermions to electrons we undo the Chern--Simons
transformation by attaching two unsmeared flux quanta to each
composite fermion, parallel to the external field direction. The
electronic problem we get is that of electrons at filling factor
$\nu=1/2$ that carry a charge smeared over a distance $Q^{-1}$, and a
total of zero magnetic flux.  This total flux, however, is made of two
smeared flux quanta anti-parallel to the external field and two
$\delta$-function flux quanta parallel to the field. Due to the
smeared flux, a moving electron exerts a transverse electric field on
all electrons within a distance $Q^{-1}$. This electric field is
proportional to the velocity of the electron, which is of order $1/m$.
Thus, the ground state of electrons experiencing this kind of
interaction is not made solely of states from the lowest Landau level,
and the effective mass of these electrons should not be expected to
renormalize to a scale determined by Coulomb electron-electron
interaction. This observation explains why the bare mass does appear
in the RPA response functions such as Eq. (\ref{kee}).

As we mentioned in previous sections, barring phase transitions, we
expect that, in the small $Q$ limit, the random phase approximation
becomes exact for response functions.
We now elaborate on that point. The RPA amounts to
calculating the response functions for a particular wavevector and
frequency ${\bf q},\omega$ by summing all diagrams in which all the
interaction lines are constrained to carry the momentum $\bf q$ and
energy $\omega$.  In other words, the RPA neglects all the diagrams
that contain interaction lines in which the momentum exchange is
summed over. Examples of both types of diagrams are shown in Fig.
2.

\begin{figure}
\label{RPAnonrpa}
  \iflongversion
  {\bf Fig2a and 2b Go Here \\}
  \else
  \centerline{\epsfig{figure=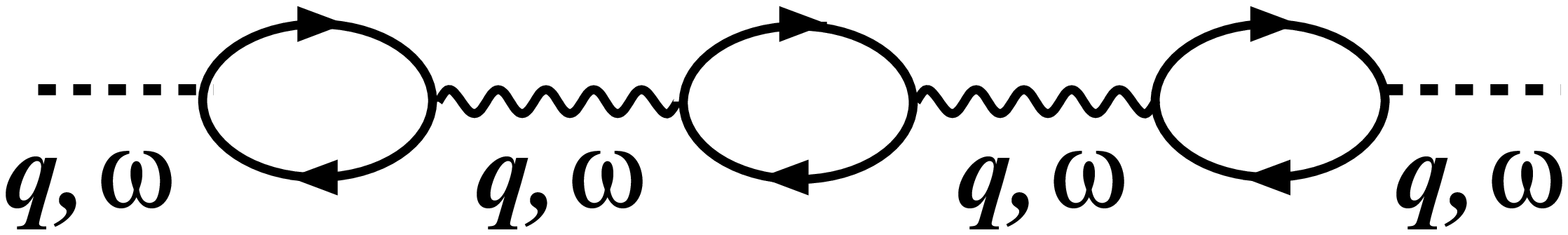,width=5.5cm,angle=270}}
  \centerline{\epsfig{figure=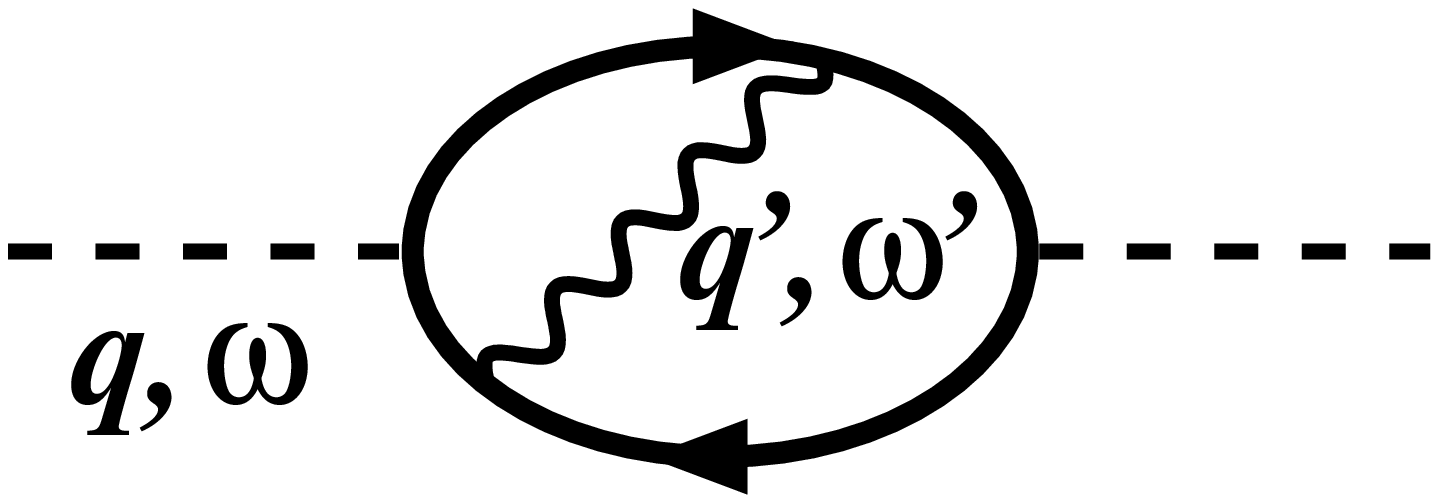,width=3.5cm,angle=270}}
  \fi
  \mycaption{{\bf Fig. 2 :} {\bf a}. (top) A diagram included in the
      RPA for the density-density response functions. The momentum
      flowing in the interaction line is restricted to be the external
      momentum $\bf q$.
      {\bf b} (bottom). A diagram neglected by RPA.  The momentum in
      the interaction line is summed over.}
\end{figure}

Consider now one of the diagrams that are neglected by RPA. The sum
over momentum exchange by the interaction is limited to $0<q\le Q$,
and in the limit of a large system, this sum may be replaced by an
integral over $0<q<Q$.  If this integral is dominated by its upper
cut-off then for small $Q$, the contribution of the diagram is at
least of order $Q^2$. To lowest order in $Q$, it may therefore be
neglected. (Note that this does not constitute a neglect of the
interaction altogether. Interaction lines appear in the RPA diagrams,
in which their momentum is not summed over).

Unfortunately, the smallness of $Q$ does not {\it always} make the
contribution of a diagram negligible, even if the momentum exchange is
summed over.  Exceptions are diagrams in which the integrand diverges
sufficiently fast at small wave vectors that the integration over
momentum exchange is dominated by an infra-red cut-off. In these cases
the result may be independent of (or weakly dependent on) $Q$. An
example of this is found in the computation of the composite fermion
effective mass.

It is well known that, using the FCS approach, one finds a divergence
in the quasiparticle effective mass at the Fermi energy
\cite{HLR,SternHalperin}, which has a logarithmic form in the case of
Coulomb interactions between electrons.  This divergence also occurs
if an upper momentum cutoff $Q$ is incorporated into the FCS
calculation.  Moreover, the coefficient of the logarithm is
independent of the value of $Q$.  However, the logarithmic
contribution only occurs for quasiparticles whose distance from the
Fermi surface is smaller than order $Q^2$, which is a very small
region of phase space when $Q$ is small.  (For the case of short range
interactions, the divergence in the effective mass is found to be
stronger than in the Coulomb case, but it only occurs for
quasiparticles whose wavevectors are closer to the Fermi surface than
a distance of order $Q^3$.)  We shall see in the next Section that 
similar singularities are found in the dipolar approach as well.

If the singularity in the quasiparticle effective mass were to carry
over to the density-density response function, we would expect that
the RPA results would become invalid at very small wavevectors, say
for $q \ll {\cal O}(Q^2)$, in the Coulomb case.  The RPA would still
be valid, and non-trivial, in the range ${\cal O}(Q^2) < q < {\cal
  O}(Q)$, so that the agreement between the FCS and dipolar approaches
found in the previous section would at least be meaningful in that
range. However, all existing analyses of these singularities at
$\nu=1/2$, by perturbative methods, renormalization group, $1/N$
expansion and bosonization find the self energy singularities to
cancel out from response functions, and the response functions to be
regular at small $q$ \cite{SternHalperin,DivergencesCancel,Marston}.
Thus, the identification of the small $Q$ limit with RPA, for response
functions, appears to be correct for arbitrarily small wavevector $q$.

\section{Fermion Green's Function and the effective mass }

In this section, we discuss the one-fermion Green's function, and the
effective mass which may be derived from it, comparing results from
the FCS and dipolar-quasiparticle approaches.  (Here, we restrict
ourselves to the case of shifted quasiparticles, as derived in Section
III using the Hamiltonian approach.)

The imaginary part of the one-fermion Green's function $G({\bf k},
\omega)$ describes the spectral resolution of the state where a
composite fermion of momentum ${\bf k }$ is instantaneously added to
or removed from the ground state.  Naturally, the Green's function
depends on the precise definition of the injected particle, and will
clearly depend on the gauge that is used, if there are any vector
potentials coupling to the fermion.  However, if the Green's function
contains a sharp pole which can be identified as arising from a
low-energy quasiparticle excitation, then the behavior of the
quasiparticle energy $\epsilon_k$ near the Fermi surface should be
well defined and should be independent of the precise definition of
the bare injected quasiparticle, as long as there is a reasonable
degree of overlap between the bare particle and the low-energy
excitation.

\subsection{Effective mass}
\label{sec:mstar}

The effective mass $m^*$ is defined by
\begin{equation}
\kf / m^*  \equiv d \epsilon_k / dk
\label{mstarH}
\end{equation}
in the limit $k \to \kf$.  At least in the case of Coulomb
interactions, it is expected that the decay rate of a quasiparticle
should be small compared to the real part of the energy, near the
Fermi surface, so that the $m^*$ is well defined in this limit.

Although the effective mass does not appear directly in the
density-density correlation function at long wavelengths and low
frequencies, it is nevertheless measurable, in principle. For example,
in a system precisely at $\nu = 1/2$, with no impurity scattering, the
specific heat at low temperatures can be related to $m^*$.  Similarly,
the behavior of $m^*$ near the Fermi surface at $\nu = 1/2$ determines
the asymptotic form of the energy gaps in the principal quantized Hall
states, $\nu = p/(2p+1)$, for $p \to \infty$\cite{HLR,SternHalperin}.

As was mentioned previously, in the FCS approach a perturbative
calculation of the composite fermion self energy leads to the
discovery that the composite fermion effective mass depends on energy,
and diverges as the energy of the composite fermion gets close to the
Fermi energy. This divergence is of the infra-red type, resulting from
the interaction of the fermion with low-energy transverse fluctuations
in the Chern-Simons gauge field, which result, in turn, from long
wavelength low- frequency fluctuations in the fermion density, due to
the constraint ${\bf \nabla}\times{\bf a}_\cs \propto \rho^e$ (Eq.
\ref{csconstraint}). The divergence is found already at the RPA level,
as a result of the contribution to the self-energy from the diagram
shown in Fig (4.a), but it has been argued to be valid exactly, at
least in the case of Coulomb interactions\cite{SternHalperin,Aim}.  A
detailed discussion of the effective-mass divergence in the FCS
approach is given in Ref.  \onlinecite{SternHalperin}.

In the Murthy-Shankar calculation, by contrast, the composite fermion
mass was found to be renormalized from its bare value to a constant
$m^*$, independent of the momentum and energy of the fermion.  The
value of $m^*$ was determined by the strength of the electron-electron
interaction and was independent of the bare mass.  However, this
result depended crucially on several aspects of the approximation used
by MS, including particularly the fact that the ultraviolet cutoff in
their theory was chosen to be precisely equal to $\kf$.  There were no
infrared divergences in their approximation.

Motivated by these findings, we now study the effective mass within
our small $Q$ model, using the formalism of Section III. With respect
to the infra-red divergence, we may envision several possible
scenarios: the effective mass divergence may be discovered as
resulting from the Hamiltonian (\ref{hd}); it may result from terms
which are of higher order in $Q$, and are therefore absent from the
Hamiltonian (\ref{hd}); or it may be absent here altogether, a result
which would raise questions about the results obtained from the FCS
approach. We find the first scenario to happen. With respect to the
ultra-violet renormalization, we may expect to find a smaller
renormalization than the one found by MS, since we take $Q$ to be
small. We indeed find this to be the case, and clarify the different
regimes at which each of the two calculations dominate.

We first note that a Hartree-Fock treatment of the Hamiltonian
(\ref{hd}) leads, in the small $Q$ limit, to a small renormalization
of the bare mass, of order $Q$. To leading order in $Q$ the
Hartree-Fock contribution to the self energy, $\Sigma_{HF}$ is
dominated by the second term in (\ref{hd}). It is described by the
diagram in Fig. (3), and is given by
\begin{equation}
\Sigma_{HF}({\bf k},E)=-\frac{1}{mn}\sum_{0<|\bf q|<Q}(1-n_F({\bf k}+{\bf 
q}))
\label{sigmahf}
\end{equation}
where $n_F$ is the Fermi occupation number.
Since $\Sigma_{HF}$ depends on wavevector only, it 
leads to a mass renormalization 
according to, 
\begin{equation}
\frac{1}{m^*}-\frac{1}{m}=\frac{1}{\kf}
\frac{\partial\Sigma_{HF}}{\partial k}=-\frac{2Q}{m\pi\kf}
\label{mass}
\end{equation}
This renormalization has a similar source as the one found by MS.

\begin{figure}
  \iflongversion
   {\bf Fig 3 Goes here \\}
  \else
  \centerline{\epsfig{figure=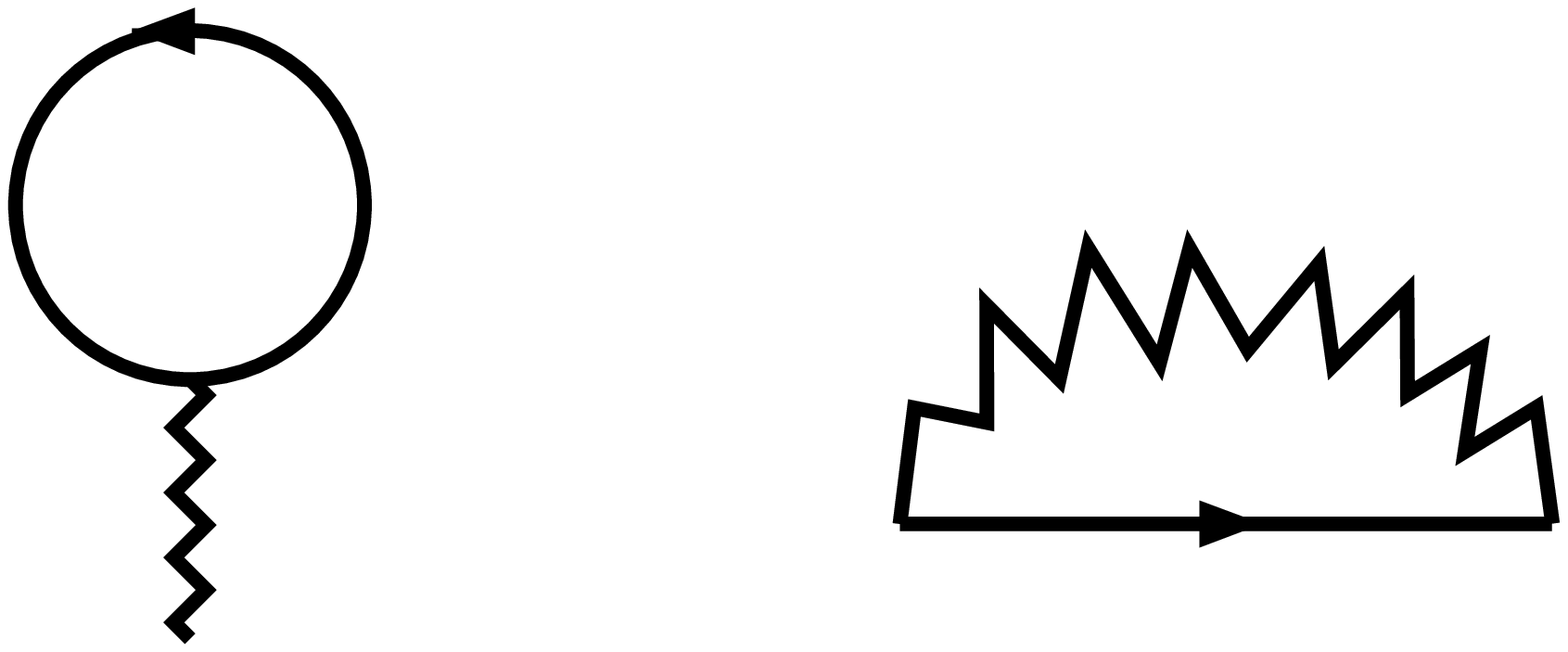,width=3.5cm,angle=270}}
  \fi
  \mycaption{{\bf Fig. 3:} The Hartree-Fock diagrams. The zig-zag
    interaction line in these diagrams represent bare ${\bf
      g}\cdot{\bf g}$ interaction (cf. Eq. (\ref{hd}). The fermionic
    lines represent bare Green functions. As usual, the Hartree
    diagram (left) does not make a contribution. The Fock diagram
    (right) leads to Eq. (\ref{sigmahf}).  }
\label{hffig}
\end{figure}

Going beyond the Hartree-Fock approximation, we replace the bare ${\bf
  g}\cdot{\bf g}$ interaction by the dressed one, which means that we
must include in the self-energy the diagram shown in Fig (4.b), as
well as the Hartree-Fock term of Fig. (3). The contribution of Fig
(4.b) leads to the same infra-red singularities as in the FCS
approach, including the same divergence of the effective mass.  The
role played in the FCS approach by the interaction with transverse
gauge field fluctuations is now played by the transverse part of the
second term in Eq. (\ref{hd}), namely the transverse part of the
fermion current-current interaction. The transverse fermion $\langle
{\bf g\bf g}\rangle$ propagator is related to the electronic density
response function (\ref{kee}), at low frequencies, by the constraint
${\bf \nabla}\times{\bf g}\propto \rho^e$ (Eq. \ref{rhoe}), which
makes its $q,\omega$ dependence exactly identical to that of the
transverse gauge field propagator in the FCS approach, and leads to
the same infra-red singularities in perturbative calculations. (We
have already seen that the electron density-density propagator is the
same, at the RPA level, in both approaches.)

The divergence of $m^*$ as a function of the energy $E$ is a
consequence of the strong dependence of the self-energy on $E$. This
dependence results from virtual processes of energy transfers $\sim E$
and momentum transfer of order $q_0\sim E^{1/3}$ for short range
interaction and $q_0\sim E^{1/2}$ for $1/r$ Coulomb interactions.  If
$E$ is larger than order $Q^3$ (or $Q^2$ in the Coulomb case), the
dependence of the self energy on $E$ weakens and the mass
renormalization from Fig (4.b) becomes small. In that regime, the
biggest contribution to the mass renormalization comes from the
Hartree-Fock term previously discussed, which gives an effective mass
which is independent of the energy or wavevector of the quasiparticle,
provided that the quasiparticle energy $\epsilon _k$ is still small
compared to $Q/ \vf$.

\begin{figure}
  \iflongversion
  {\bf Fig 4 goes here \\} 
  \else
  \centerline{\epsfig{figure=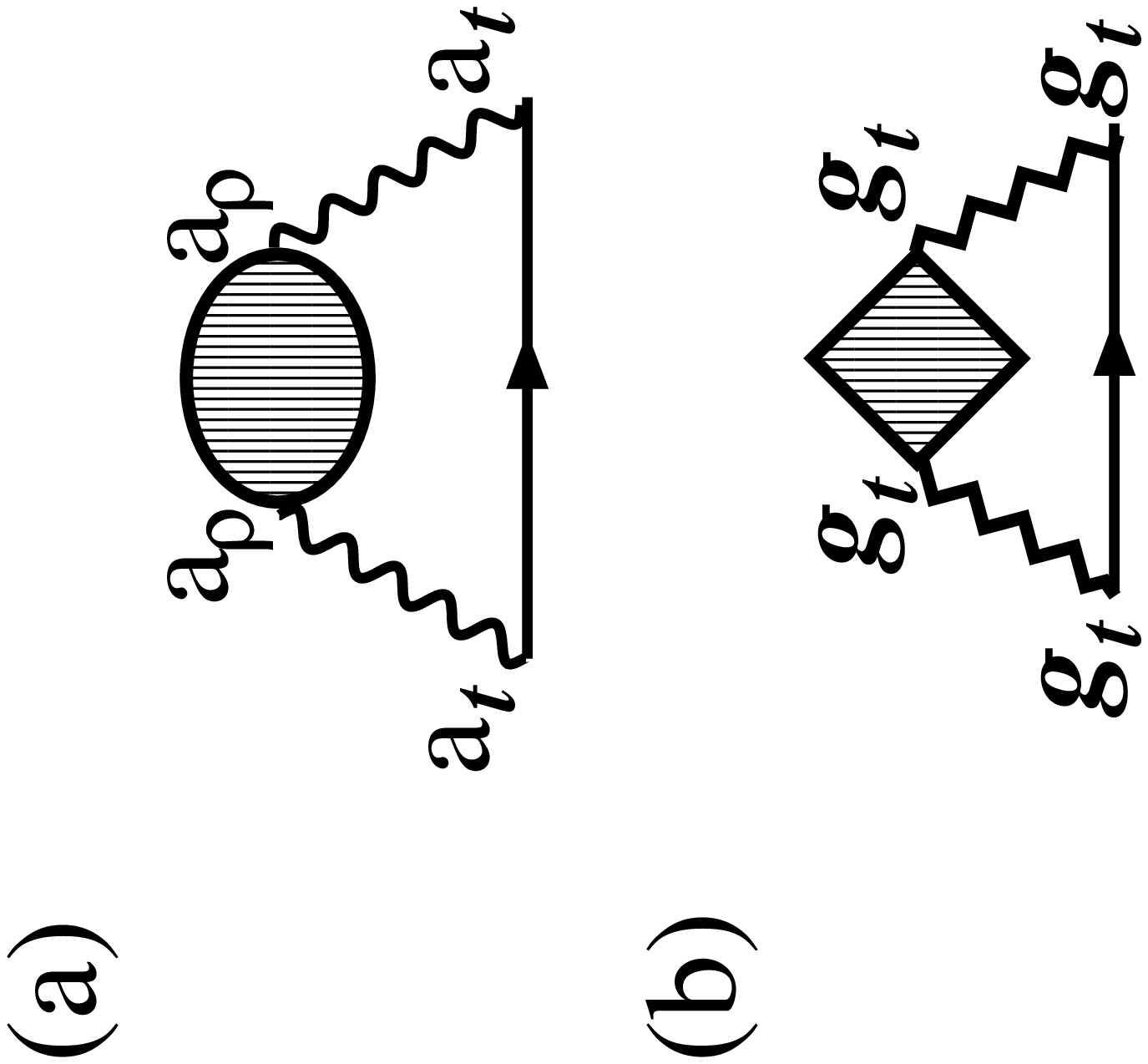,width=3.5cm,angle=270}}
  \fi
  \mycaption{ {\bf Fig 4 : (a)}. The fermion self energy in the FCS
    approach. Here, a wiggly line is a Chern--Simons interaction, and
    the elliptic full bubble represent the electronic density-density
    correlation function. The two wiggly lines combined with the full
    bubble form the transverse gauge field propagator $\langle a_t
    a_t\rangle$ due to the Chern--Simons constraint,
    (\ref{csconstraint}).  {\bf (b)}. The fermion self energy in the
    shifted quasiparticle picture is the sum of this diagram and the Fock
    diagram in Fig. 3. A zig-zag line is a bare
    transverse momentum interaction, while the full diamond shaped
    bubble is the $\langle g_t g_t\rangle$ correlation function. At
    low frequencies, this correlation function is related to
    electronic density-density response function (the elliptic bubble
    of (a)) by the relation (\ref{udens}). At high frequencies the
    $g_t g_t$ interaction is dominated by the bare term in Fig. 3,
    as explained in the text.  }
\label{xx}
\end{figure}

In the FCS approach, one can obtain a contribution to the effective mass 
identical to that of the Hartree-Fock diagram in the present approach by 
including in the self-energy of Fig (4.a) the contribution to the 
transverse gauge fluctuations arising from the high-frequency Kohn mode.  
Since this contribution does not lead to any singularity at low energies, 
it has generally been ignored in the FCS literature.

We believe that the general state of affairs, which we find to hold
for small $Q$, holds also for the physical case where no upper cut-off
$Q$ exists. For low enough energies, the effective mass is determined
by singular infra-red contributions, and diverges close to the Fermi
surface.  The leading singularity in the effective mass is independent
of $Q$, at least in the case of Coulomb interactions. This is another
manifestation of the statement made in Ref. (\onlinecite{SternHalperin}):
the leading singularity in $m^*$ is captured exactly by the
perturbative calculation.  For high enough energies, on the other
hand, the effective mass is determined by the short distance behavior
of the interaction. In the physical case, unlike the small $Q$ limit,
the high energy effective mass is renormalized from the bare mass
scale to a scale determined by electron-electron interaction.

The fact that the low-energy divergence of the effective mass depends
on the range of electron-electron interaction leads us to express a
word of caution regarding numerical calculations of the effective
mass.  The singular $q,\omega$ dependence of Eq. (\ref{kee}), in all
its derivations, is obtained by summing infinitely many terms (the RPA
geometric series), where the $n$'th term represents the amplitude for
excitation of $n$ particle-hole pairs. Numerical calculations based on
trial wavefunctions often calculate the effective mass by comparing
the energy of a ground state with the energy of an excited state with
{\it one} excited particle hole-pair.  Although Jain type trial wave
functions used in these calculations are not the wave functions that
emerge from our analysis, they are not unrelated (see the discussion
by MS\cite{Shankar}). Moreover, the trial wave functions used in
numerical calculations are constructed in a way which is independent
of the range of electron-electron interaction. Thus, we believe that
calculations done using these wavefunctions are very useful in
understanding the behavior of the effective mass at high energies, but
are not able to discover the infra-red divergence at low energies,
which depends strongly on the range of electron-electron interaction
and results from many particle-hole pairs being excited.

\subsection{Quasiparticle-weight in the fermion Green's function.}

As has been noted in the literature,  the amplitude $z$ of the
single-quasiparticle contribution to the Green's function for the
bare composite fermion of the FCS approach actually vanishes in the 
thermodynamic limit and at low energies. In fact there are two separate
reasons for this.

The first point, noted by HLR\cite{HLR}, is that, in the Coulomb
gauge, the operator which inserts a bare composite fermion at some
instant of time will excite a number $n_m$ of long-wavelength
magnetoplasma modes, whose mean value diverges logarithmically with
the size of the system.  This means that the probability of winding up
in the ground state of the magnetoplasma modes falls off as a power of
the area of the system, and the quasiparticle amplitude $z$ must fall
off accordingly.  Although this effect is an infrared divergence in
the sense that it depends on the size of the system, it is not
sensitive to the wavevector or energy of the quasiparticle, since the
responsible fluctuations are at very high energy.  In fact, the
problem occurs equally well for a fractional quantized Hall state,
with an energy gap, as it does for $\nu=1/2$.  Within the HLR
formalism, this diverging contribution arises from interactions of the
bare composite fermion with fluctuations in the Chern-Simons scalar
potential associated with the magnetoplasmon modes.

The diverging renormalization of $z$ due to magnetoplasmon modes does not 
occur for the dipolar fermions we consider here, because the fermions are 
decoupled from the oscillators at long wavelengths.  This makes sense, 
because  the fermion Green's function here describes the propagation of 
an added composite fermion after all states with non-zero oscillator 
occupation have been eliminated. 

The second type of divergent renormalization of the quasiparticle
weight $z$, encountered in the FCS approach, is an infrared
divergence, resulting from interactions with fluctuations in the
transverse Chern-Simons field due to the low-frequency density mode.
This renormalization of $z$ arises from the same frequency-dependent
contribution to the self-energy as is responsible for the infra-red
singularity in the effective mass.  This renormalization is present
equally well for the dipolar fermions considered in the present
section.  The quasiparticle weight is thus predicted to vanish as a
power of the quasiparticle energy $E$, in the case of short-range
interactions, and to vanish as $ 1/ \log E$ in the case of Coulomb
interactions.

The fact that the quasiparticle weight in the one-fermion Green's
function vanishes at low energies does not lead to any major effects
in the density-density response, as has been previously noted, because
all the diverging contributions are cancelled in this
case\cite{DivergencesCancel}.  The diverging contribution of the
long-wavelength magnetoplasmons which led to a vanishing of the
quasiparticle weight in the one-fermion Green's function of the FCS
theory is absent in the density response function because of the
opposite signs of the interaction for the particle and the hole.
(This has been discussed explicitly in Ref. \onlinecite{KivelsonLee})

Finally, we must note that in calculating the one-fermion Green's
function for our dipolar quasiparticles, we have implicitly assumed
that the initial state has zero total momentum, as well as vanishing
amplitude for all the unphysical modes of the longitudinal momentum
density, which occur at zero-frequency and non-zero $\bf q$. (This is
similar to fixing the gauge when calculating the Green's function for
a charged particle.) Clearly, if we considered the physically
equivalent state obtained by adding a constant momentum ${\bf K}$ to
every fermion, we would find a Green's function where the Fermi
surface was itself displaced by the constant momentum.  If one
averaged the Green's function over a set of physically equivalent
initial states with different values of ${\bf K}$ one would lose all
useful information, as would also happen if one averaged over initial
states with non-zero amplitudes in unphysical zero frequency and
non-zero $\bf q$ modes.

\vspace*{80pt}

\section{Large Q: Fermi liquid picture}
\subsection{General conjectures}
In the previous sections we discussed in detail a model in which the
random phase approximation is exact. 
 In this section, we conjecture the form of a Fermi liquid theory
for the dipolar quasiparticles which describes the electronic response
functions for the physical case, in which there is no momentum cut-off 
$Q$, and discuss what features of the RPA
are expected to remain valid in the physical case.

As we saw in previous sections,   there are various different
formulations of the small $Q$ limit, such as the electron-centered and
shifted quasiparticles. Naturally, this ambiguity
carries over to the physical case. Below, we first enumerate four
assumptions which we believe to be common to all Fermi-liquid
descriptions of the $\nu=1/2$ state 
in terms of dipolar quasiparticles. We then consider the
static limit and show that these assumptions are already sufficient to
establish that the electronic system is compressible. In a second
step, we specify to electron-centered quasiparticles and discuss
 the  dynamic density-density response function.

We conjecture that any Fermi-liquid theory of the $\nu=1/2$ state in 
terms of dipolar quasiparticles
will satisfy the following four assumptions:

\begin{enumerate}
\item \label{dipoleconj}
The electronic density is to be described in terms of low-energy
  quasi-particles and high-energy magnetoplasmons.  The low energy
  quasi-particles are neutral, and carry an electric dipole moment
  proportional and perpendicular to their momentum. Thus, in the limit
  of small ${\bf q}$ and $\omega$, the electronic density $\rho^e$ and
  the quasi-particle momentum density ${\bf g}$ are related by
\begin{equation}
\rho^e=\frac{i{\bf q}\times{\bf g}}{2\pi{\tilde\phi}n}.
\label{assum1}
\end{equation}
This relation was found to hold in the RPA approach for both
electron-centered and shifted quasiparticles. 

The relation (\ref{assum1}) is valid precisely at $\nu=1/2$. If an 
infinitesimal driving force is applied to the system by means of a vector 
potential $\bf A_{{\bf q},\omega}$ then Eq. (\ref{assum1}) has to be 
modified to, 
\begin{equation}
\rho^e=\frac{i{\bf q}\times{\bf g}}{2\pi{\tilde\phi}n}+
       \frac{i{\bf q}\times{\bf A}}{2\pi{\tilde\phi}}.
\label{assum1prime}
\end{equation}
This means that if the electron-density $\rho^e$ tracks the magnetic field,
so that the system is everywhere  {\it locally} at $\nu=1/2$, we still have
${\bf q}\times{\bf g}=0$.

\item \label{fltconj} The low energy quasi-particles form a Fermi
  liquid characterized by response functions that are regular in the
  small ${\bf q},\omega$ limit. Moreover, we assume that this liquid
  is compressible.
  
\item \label{Kinvconj} The Fermi liquid theory satisfies
  $K$-invariance, i.e., the quasi-particles' energy is invariant to a
  shift of the Fermi surface by a constant $\bf K$. Moreover, the
  invariance holds even for a position dependent boost such as ${\bf
    K}={\bf K_0} \cos ({\bf q}\cdot{\bf r})$, with ${\bf K}_0||{\bf
    q}$.  Put in different words, this Fermi liquid possesses
  zero-energy excitation modes at small non-zero values of $q$,
  associated with static longitudinal fluctuations in the momentum
  density ${\bf g}$.  This assumption carries over from the RPA
  results.  The zero-energy modes do not appear in any physical
  observables.
  
\item \label{rgconj} The quasi-particle density $\rho$ is related to
  the electronic density $\rho^e$ (or, alternatively, to the
  quasi-particle momentum density ${\bf g}$ due to Eq.\ 
  (\ref{assum1})) in a non-singular manner in the limit $q\rightarrow
  0$. For small $Q$, we found $\rho({\bf q}) \propto \rho^e({\bf q})$
  with the proportionality constant varying between different
  descriptions.  We assume that the proportionality relation continues
  to hold for large $Q$, though the constant of proportionality may be
  modified in some of the descriptions.
 
\end{enumerate}
  
In our discussions we ignore the possibility of infrared divergences
in the quasiparticle effective mass $m^*$, and we implicitly assume
that the decay rate of a quasiparticle is small compared to its
energy.  These assumptions should be literally correct if the
electron-electron interaction is longer range than $1/r$. The
arguments go through with relatively minor modifications in the case
of Coulomb interactions, where there is a weak logarithmic divergence
of the effective mass, and the quasiparticle decay rate is
asymptotically small compared to the energy \cite{SternHalperin}.  
It is not clear how much
of the Fermi-liquid description can be retained for short-ranged
interactions, however, where the quasiparticle decay rate is predicted
to be proportional to the energy, as one approaches the Fermi surface.

\subsection{Static response}
Making the assumptions above, we now show that the $\nu=1/2$ state is
compressible.  An electronic system is compressible if the energy cost
involved in producing a (static) modulation $\delta\rho^e({\bf q})$ of
the electronic density is finite for $q\rightarrow 0$.  (Here, we
exclude the direct Coulomb energy of the density fluctuation, which,
of course, diverges for $q \rightarrow 0$.)  In the present case, due
to assumptions (\ref{dipoleconj}) and (\ref{rgconj}), a modulation
$\delta\rho^e$ in the charge density is associated with both a
quasiparticle momentum density $\delta {\bf g}$ (of order $q^{-1}$)
and a quasiparticle density modulation $\delta\rho$ (of order $q^0$).
The energy cost associated with a small $\delta\rho^e$ can then be
written as,
\begin{equation}
  \delta E(\delta\rho^e)=\frac{1}{2}{\cal A}(\delta {\bf g})^2+
  {\cal B}{\delta g}\delta\rho+\frac{1}{2}{\cal C}(\delta\rho)^2
\label{energy}
\end{equation}
In this expression, ${\cal A}$ should vanish at least as fast as $q^2$
for small $q$, as a consequence of ${\bf K}$ invariance and the
assumption (\ref{fltconj}).  The energy cost associated with the first
term of (\ref{energy}) is then finite.  In the second term, $\cal B$
must be at least of order $q$ as a consequence of assumption
(\ref{fltconj}), and thus the energy cost associated with this term is
finite.  Finally, the assumption of a compressible quasi-particle
Fermi liquid requires $\cal C$ to be a constant in the limit
$q\rightarrow 0$.  Combining all terms in (\ref{energy}) we find
$\delta E({\delta\rho^e})$ to be finite.

\subsection{Dynamical response}
Generally, dynamical response in Fermi liquid theory is analyzed by
means of a Boltzmann equation, which describes the non-equilibrium
state of the liquid in terms of the function $\delta n({\bf k},{\bf
  r},t)$, the deviation of the quasiparticle distribution function
from the equilibrium Fermi-Dirac function.  At zero temperature one
may write $\delta n({\bf k},{\bf
  r},t)=(2\pi/m^*)\delta(\epsilon_k-\mu)\nu(\theta , {\bf r}, t)$
(where $\theta$ is the angle of the vector $\bf k$ with the $x$-axis),
and expand $\nu(\theta,{\bf r},t)=\sum_l\nu_l \exp(il\theta)$.  The
quasi-particle density is then $\rho({\bf r},t)=\nu_0({\bf r},t)$,
while the two components of the kinetic momentum density are
\begin{eqnarray}
g_{x}({\bf r},t)+n_0 A_x({\bf r},t)&\equiv&\sum_{\bf k}{\bf k}_x
\,\delta n({\bf k},{\bf r},t) \nonumber \\ &=&\frac{1}{2}\kf(\nu_1({\bf r},t)
+\nu_{-1}({\bf r},t))\nonumber \\
g_{y}({\bf r},t)+n_0 A_y({\bf r},t)&\equiv&\sum_{\bf k}{\bf k}_y
\,\delta n({\bf k},{\bf r},t) \nonumber \\
&=&\frac{i}{2}\kf(\nu_1({\bf r},t) - \nu_{-1}
({\bf r},t))
\label{fltg}
\end{eqnarray} 
where ${\bf A}({\bf r},t)$ is an infinitesimal externally applied
driving vector potential. Note that in the presence of ${\bf A}({\bf
  r},t)$ the vector ${\bf k}$ is the kinetic momentum, ${\bf k} = {\bf
  p} + {\bf A}$, rather than the canonical momentum ${\bf p}$.  The
energy $\epsilon_k$ is thus unaffected by the vector potential and the
equilibrium Fermi sphere remains centered at $k=0$.

The kinetic momentum density is distinguished in Fermi liquid theory
from the quasi-particle current, which is expressed in terms of
$\nu_{\pm 1}$ and the (dimensionless) Landau parameters $F_{\pm 1}$
as,
\begin{equation}
\begin{array}{ccc}
j_{x}&=&\frac{1}{2}\vf^* ((1+F_1)\nu_1+(1+F_{-1})\nu_{-1})\nonumber\\
j_{y}&=&\frac{i}{2}\vf^*((1+F_1)\nu_1-(1+F_{-1})\nu_{-1})
\label{jflt}
\end{array}
\end{equation} 

It is convenient to work in momentum-frequency representation, so from now on, 
$\nu(\theta)$ and $\nu_l$ will be implicitly assumed to depend on ${\bf q}$ and 
$\omega$. The
Boltzmann equation for $\nu(\theta)$ in the presence of an electric
field ${\bf E}\propto e^{i{\bf q}\cdot{\bf r}-i\omega t}$ is,  
\begin{equation}
   -i\omega\nu(\theta)+iv_F^*q\cos\theta[\nu(\theta)
     +\delta\nu(\theta)]
        =-{\kf\over2\pi}{\bf E} \cdot {\bf \hat n}(\theta)
\label{Boltseq}
\end{equation}   
with ${\bf \hat n}(\theta)=(\cos\theta,\sin\theta)$ and
$\delta\nu(\theta)=\int(d\theta^\prime/2\pi)
F(\theta-\theta^\prime)\nu(\theta^\prime)$.  The 
angular Fourier components of the quasi-particle 
interaction function  $F(\theta-\theta')$ are the Landau parameters. 
The field ${\bf E}$ is the self-consistent electric field, 
including both the external probing field and the field
produced by the long-range Coulomb potential arising from the induced
inhomogeneities in the electron density.

Eq. (\ref{Boltseq}) easily yields a continuity equation
$\omega\rho={\bf q}\cdot{\bf j}$.  In our case, however, the
quasi-particle density $\rho$ and current $j$ may be related to the
electronic density $\rho^e$ and current $j^e$ in a non-trivial way.
In fact, this relation depends on the particular description we
choose, and varies between Section (\ref{sec:Lagrange}) and Section
(\ref{sec:Hamilton}).  To be specific, we now assume that we may
construct a Fermi liquid description in which the quasi-particles are
electron-centered, as in Section (\ref{sec:Lagrange}), i.e.,
\begin{equation}
      \rho^e({\bf q})=\rho({\bf q}).
\label{assum5}
\end{equation}
By the continuity equation, Eq. (\ref{assum5}) fixes the longitudinal 
component of the electronic current to be the projection of Eq. (\ref{jflt})
onto the direction of ${\bf q}$. Furthermore, Eq. (\ref{assum1prime}), when 
combined with Eqs. (\ref{jflt}) and (\ref{assum5}), allows us to identify, 
\begin{equation}
   F_{\pm 1}=-1\pm{m^*\over n_0}{\omega\over2\pi\tilde\phi}.
\label{FLTparameter}
\end{equation}
This expression for $F_{\pm1}$ is reminiscent of the results of
Section (\ref{sec:Lagrange}). In the static limit $\omega=0$ we have
$F_{\pm1}=-1$.  Generally, the cost in the total energy ($E - \mu N$)
associated with a spatially slowly varying infinitesimal static
deformation $\nu(\theta)$ of the Fermi sea is proportional to $\sum_l
(1 + F_l) |\nu_l|^2$.  A small uniform boost of the Fermi sea by
momentum $\vec K$ is described by $\nu_{{\pm}1}\sim\frac{1}{2} (K_x
\pm i K_y)$.  {\it Thus, $F_{\pm1}=-1$ eliminates any energy cost
  associated with this uniform boost of the Fermi sea in the
  zero-frequency limit.}  This is precisely the statement of ${\bf K}$
invariance. Note that in this description $K$-invariance is not an
independent assumption but rather results from Eqs. (\ref{assum1}) and
(\ref{assum5}).  At non-zero $\omega$ Eq. (\ref{FLTparameter}) has the
unusual feature of being dependent on the frequency. While the
frequency dependence of $F_1$ is hard to interpret in the context of
an energy functional, it can be quite naturally included in a
Boltzmann equation.

We now turn to use the Boltzmann equation for studying the current
${\bf j}$ induced, in linear response, by an electric field ${\bf E}$.
Before doing so, we make three comments of caution. First, we note
that the considerations above, which identified $j^e_l$ with $j_l$, do
not identify $j^e_t$ with $j_t$. In fact, such an identification does
not hold even in the small $Q$ limit (cf. Eq. \ref{dipole-current}).
Consequently, we are able to draw conclusions only with regard to the
longitudinal electronic current response. Second, we note that since
we are not able to calculate the Landau parameters $F_l$ for $l\ne\pm
1$, the information we may obtain from the Boltzmann equation is
limited. And third, as usual, the semi-classical Boltzmann equation
does not properly account for the quasi-particle Landau diamagnetism,
which, as we saw in Secs. (\ref{sec:Lagrange}) and
(\ref{sec:Hamilton}), affects the electrons' compressibility. Its
effect has to be included in the equation by hand.

As explained in detail in Refs. \onlinecite{SimonReview} and
\onlinecite{SimonHalperin}, generally the effect of $F_{\pm 1}$ on
response functions can be understood by the following procedure: we
write the Boltzmann equation as
\begin{eqnarray}
   -i\omega\nu(\theta)&+&iv_F^*q\cos\theta[\nu(\theta)
     +\int(d\theta^\prime/2\pi) 
{\tilde F}(\theta-\theta^\prime)\nu(\theta^\prime) ] \nonumber
        \\ &=&-{\kf \over2\pi}{(\bf E+E^{\rm eff})\cdot\hat n}(\theta).
\label{Boltseq2}
\end{eqnarray}   
where
\begin{equation}
 {\bf E}^{\rm eff}=-{2\pi i\omega\over \kf^2}\left[(F_1+F_{-1}){\bf g}
     +i(F_1-F_{-1}){\bf\hat z}\times{\bf g}\right],
\end{equation}
and ${\tilde F}(\theta)=F(\theta)-F_1e^{i\theta}-F_{-1}e^{-i\theta}$. 
(The Landau parameters ${\tilde F}_l$ 
corresponding to ${\tilde F}(\theta)$ are those corresponding 
to $F(\theta)$, except for the case $l=\pm 1$, 
in which ${\tilde F}_{\pm 1}=0$). 
For our particular values of $F_{\pm 1}$ (Eq. (\ref{FLTparameter})) and 
given Eqs. (\ref{fltg}) and (\ref{jflt}), we get
\begin{equation}
 {\bf E}^{\rm eff}=-2\pi\tilde\phi{\bf \hat z}\times{\bf j}
     -{i\omega\over n_0/m^*}{\bf j}.
\label{aeeff}
\end{equation}
The second term in (\ref{aeeff}) is ${\cal O}(\omega)$ smaller than 
the first, and can therefore be neglected. 

If we define a conductivity tensor $\sigma({\bf q}, \omega)$ by
${\bf j}=\sigma {\bf E}$, and a ``quasiparticle
conductivity tensor'' $\sigma^*({\bf q},\omega) $ by ${\bf j}=\sigma^*
({\bf E}+{\bf E^{\rm eff}})$, then the two quantities are related by
\begin{equation}
   \sigma^{-1}=(\sigma^*)^{-1}+2\pi\tilde\phi\left({0\atop
       1}{-1\atop 0}\right).
\label{fltres}
\end{equation}
The quantity $\sigma^*({\bf q}, \omega) $, in turn, is the 
linear response function which gives
the relation between induced current and the electric field for a
conventional  Fermi liquid, with 
effective mass $m^*$ and Landau parameters ${\tilde F}_l$.

Eq. (\ref{fltres}) also clarifies the relation between the Fermi liquid 
picture we develop here from the dipolar approach and the Fermi 
liquid picture developed in the FCS approach. In the latter one gets 
Eq. (\ref{fltres}) with one more term on the right hand side, given by 
$\frac{i\omega}{n_0}(m^*-m)$. This term does not affect the response at 
low frequency, but is essential for getting the correct electronic 
response at $\omega\approx \omega_c$, particularly Kohn's mode. Thus, 
our dipolar picture, which we confined to low frequency, coincides with 
the FCS approach in that limit\cite{SimonReview,SimonHalperin}. 

As we emphasized before, only one component of $\sigma({\bf q},\omega)$,
 namely $\sigma_{ll} (q,\omega)$, can be ascribed a physical meaning. 
The electronic density-density response function 
is related to that term by 
\begin{equation}
   K^e_{\rho\rho}({\bf q},\omega)
      ={1\over v(q) -i\omega / q^2 \sigma_{ll} }.
      \label{keeBH}
\end{equation}
In the limit $\omega,q\rightarrow 0$ and $\frac{\omega}{q}\rightarrow
0$, this takes the form,
\begin{equation}
   K^e_{\rho\rho}({\bf q},\omega)
      ={1\over v(q)+{\tilde {\cal C}}
      -i(2\pi\tilde\phi)^2{2n_0\omega\over \kf q^3}}.
\label{keeflt}
\end{equation}
where $ {\tilde {\cal C}}$ is the inverse compressibility of the  Fermi liquid, 
\begin{equation}
 {\tilde{\cal C}} = 2 \pi (1+F_0)/ m^*.
\label{calC}
\end{equation}

The functional dependence of (\ref{keeflt}) on ${\bf q},\omega$ is the
same as that of the small $Q$ result, Eq. (\ref{kee}). The constant
$\tilde{\cal C}$, however, is modified. We are not able to calculate
the value of $\tilde{\cal C}$ in terms of the original parameters of
the electronic problem. Moreover, it is not clear to us whether there
is a unique way of relating $\tilde{\cal C}$, which is an electronic
property, to Fermi liquid properties of the quasi--particles. Due to
the constraint (\ref{assum1prime}), for a longitudinal $\bf A$ a
quasi-particle density excitation must involve an excitation of a
transverse ${\bf g}$. In the absence of the constraint, the dynamics
of a density excitation involves $F_0$ and is independent of the
diamagnetic susceptibility $\chi$, while the dynamics of a $g_t$
excitation involves $\chi$ and is independent of $F_0$. Thus, $F_0$
and $\chi$ are two independent quantities. Here, with the two
excitations correlated, it is not clear whether $F_0$ and $\chi$ are
uniquely and independently defined.

A further complication arises because the singular nature of the
dipolar system, where $1+F_{\pm 1}=0$, means that corrections to the
usual Fermi liquid theory can change the value of ${\tilde{\cal C}}$,
even within a calculation based on the Boltzmann equation.  In
particular, at low frequencies, in the long wavelength limit, it may
not be not sufficient to use a Landau interaction function
$F(\theta-\theta^\prime)$ appropriate to $q=0$; we find that
corrections to $F$ that are proportional to $q$ can affect the value
of $\tilde{\cal C}$ in the limit $q \to 0$.

It should be noted that the value of $\tilde{\cal C}$ does not affect
the leading behavior of the density response function at long
wavelengths, for any frequency, in the case of Coulomb interactions,
because the constant $\tilde{\cal C}$ is small compared to the
interaction term $v(q)$.  Even in the case of short-range
interactions, the value of $\tilde{\cal C}$ is unimportant for the
response function as long as $ \omega \gg \vf q^3/ \kf^2$.

The Fermi liquid picture we obtain shares some similarities with the
Fermi liquid obtained recently by Read (Ref. \onlinecite{ReadNew}) for
bosons at $\nu=1$ constrained to the lowest Landau level. In
particular, the identification of $F_1=-1$ ($K$--invariance) and the
linear response function (\ref{keeflt}) are obtained also in Read's
approach. This similarity indicates that some of the features obtained
in a calculation constrained to the lowest Landau level are, in fact,
unchanged when this constraint is removed. In particular, our study
concludes that for Coulomb interactions the long wavelength limit of
(\ref{keeflt}) (where $\frac{1}{v(q)}\gg {\tilde{\cal C}}$) is valid
independent of whether the electrons are confined to the lowest Landau
level. Moreover, the diverging effective mass and the energy gaps at
filling factors $p/(2p+1)$ with $p\rightarrow \infty$ are fully
determined by the long wavelength limit of (\ref{keeflt}), and are
found to depend only on the scale of electron--electron interaction
\cite{HLR,SternHalperin}.  Consequently, we conclude that, in the case
of Coulomb interaction, for any ratio of the interaction energy to the
cyclotron energy at which electrons form fractional quantized Hall
states at $\nu=p/(2p+1)$ the energy gaps correponding to these states
are solely determined by the interaction energy in the limit
$p\rightarrow\infty$. In practice, since this statement results from
the weak, logarithmic, divergence of the effective mass, the limit
$p\rightarrow\infty$ may be realized only for extremely large values
of $p$.

\section{Conclusions}

When analyzing the role of electron-electron interaction in the
physics of electrons in zero magnetic field, the common strategy
treats the interaction as a perturbation. The interaction may then be
handled by means of a Hartree-Fock approximation, an RPA, or a Fermi
liquid approach. These schemes attempt to find (approximately) what
are the low energy excitations (quasi-particles and quasi-holes) of
the problem, and using these excitations, what is the electronic
response of the system to driving forces.

This strategy cannot be directly applied to the case of a half filled Landau 
level, since in the absence of the interaction the ground state is vastly 
degenerate. The Leinaas-Myrheim-Chern-Simons transformation, 
which attaches ${\tilde\phi}=2$ flux 
quanta to each electron, opens the way to a perturbative calculation in which 
the unperturbed ground state  is non-degenerate. 
However, this transformation, 
which transforms electrons into CS fermions, generates a new interaction, the 
Chern-Simons interaction, whose coupling constant, $\tilde\phi$, is not small.

An important feature of the transformed problem is that the singular nature of
the interaction for $q \to 0$ leads to a strong coupling between the CS
fermions and the high-energy collective modes (the magnetoplasmons) at long
wave lengths.  Due to this strong interaction, the low energy excitations of
the transformed problem, the dressed composite fermions or quasi-particles,
are very different from the bare Chern-Simons fermions.  

There are some strong similarities between the fermion-Chern-Simons
system and the ordinary three-dimensional electron system, where the
Coulomb interaction gives rise to plasma modes whose frequency remains
finite at $q=0$, and where coupling to the plasma modes gives rise to
complete screening of the charge of the quasiparticle excitations at
low energies.  As in the case of the 3D electron system, there are
several ways of approaching the FCS system.  The conventional FCS
approach treats the CS interaction and the Coulomb interactions as
though they were weak perturbations, taking care of the most singular
aspects of the CS interaction by using (at least) the RPA to calculate
the long-wavelength density and current response functions of the
electron system.  An important result is the density--density response
function in the limit of small $q,\omega, \frac{\omega}{q}$. This
function has a unique $q,\omega$ dependence, which is calculated
explicitly for small $\tilde\phi$ (giving Eq. (\ref{kee})) and is
argued to be correct for the physical value of $\tilde\phi$. In
particular it predicts the $\nu=1/2$ state to be compressible.

The conventional FCS approach can also be used to extract properties of the
low-energy quasiparticles from the Green's function of the bare CS fermion. 
Although the overlap with the propagator for the low-energy quasiparticles is
vanishingly small in the limit of an infinite system, one nevertheless gains
insight into the quasiparticle effective mass $m^*$ from the structure of the
CS fermion self-energy, learning for example that there is a logarithmic
infrared divergence in case of Coulombic electron-electron
interactions.\cite{HLR,SternHalperin,DivergencesCancel,Marston,Aim}

The conventional FCS approach is analogous to the perturbative
approach most commonly employed in the study of three-dimensional
electron systems.  An alternate approach, suggested by
MS\cite{Shankar} and motivated by an earlier work of Read\cite{NickRead},
is analogous to the Bohm-Pines\cite{Bohm} treatment of the 3D electron
system.  MS fix $\tilde\phi$ at its physical value, and apply a
unitary transformation in order to produce low energy quasi-particles
which are decoupled, at long wavelengths, from the high-energy
magnetoplasmon modes.  This approach may seem intuitively more
appealing than the conventional FCS approach, since it directly brings
to light the dipolar nature of the quasi-particles, and thus naturally
explains why they move in straight lines in a strong magnetic field.
Beyond offering a different intuition, however, the MS approach
suggested also the possibility that the density-density response
function calculated by the FCS approach might be incorrect.
Particularly, it suggested the possibility that the $\nu=1/2$ state is
incompressible, as a result of the weak coupling of dipolar particles
to electromagnetic fields.

Our original goal in the present work was to settle the seeming contradiction 
between the two 
approaches, and to show that Eq. (\ref{kee}) is indeed the correct 
density--density response function for electrons at $\nu=1/2$. Our 
starting point was the Hamiltonian (\ref{ham}). We followed MS 
in using the temporal (Weyl) gauge, in which the Hamiltonian contains two sets 
of dynamical degrees of freedom, the CS fermions and the CS gauge field. 
Our first step was to formulate a simplified model which can be analyzed 
systematically by means  of an expansion in a small parameter, and apply both 
approaches to study that model. In this model, which we termed the  
small $Q$ model,  the charge and flux quanta carried by the Chern--Simons 
fermion are smeared over an area $\sim Q^{-2}$ around the fermion, and $Q$ is 
the small parameter. For the FCS perturbative analysis, a small $Q$ limit is 
equivalent to  a small $\tilde\phi$ limit, and can be  treated by the RPA. 
Having observed that, we carried out two different analyses of 
the small--$Q$ model in 
terms of dipolar quasi-particles, attempting to understand how, despite their 
weak coupling to electromagnetic fields, the dipolar quasi--particles form a 
compressible state. 

In the first analysis, described in Sec.  (\ref{sec:Lagrange}), we used a
Lagrangian formulation, and integrated out the high energy magneto-plasmon
modes, obtaining an  effective low energy action (\ref{low-e-action}) for
the Chern-Simons fermions. This approach, while close to the FCS calculation,
brought to light the dipolar nature of the CS fermions in the temporal gauge,
as was seen by the relation (\ref{dipole-current}) between the quasi-particle
momentum and the electronic current. Moreover, the action (\ref{low-e-action})
also revealed the important difference between   a gas of free fermions
carrying a dipole moment and the dipoles we have at hand. This difference is
the $K$--invariance of the present problem, the invariance of the energy of
the fermionic system to a shift of the Fermi sphere in momentum space. In the
language of Sec. (\ref{sec:Lagrange}), $K$--invariance is guaranteed by the
current--current interaction in (\ref{low-e-action}), which  compensates the
kinetic energy cost involved in a momentum space displacement of the
quasi-particle's Fermi sphere and makes the associated total energy cost
vanish.  As we saw first in Sec. (\ref{sec:Lagrange}), it is the combination
of $K$--invariance and the way the dipole moment is related to the
quasi-particle's momentum that makes the $\nu=1/2$ state compressible, and
leads to the density-density response function (\ref{kee}), which is precisely
the one obtained in the FCS approach.  
 
In the second analysis of the small--$Q$ limit, carried out in section
(\ref{sec:Hamilton}), we performed the unitary transformation
suggested by MS to decouple the low energy quasi-particles (the
fermions) from the high energy magnetoplasmons (the gauge field),
working systematically to lowest order in $Q$. We found the
transformed Hamiltonian (\ref{hd}), in which the two degrees of
freedom decouple in the long wavelength limit. We then identified the
way the electronic density is expressed in terms of the transformed
variables, and calculated the electronic density--density response
function, obtaining again Eq.  (\ref{kee}). Although the low-energy
quasiparticles do not interact with the transformed Chern-Simons gauge
field at long wavelengths, they do retain strong momentum-dependent
interactions which must be properly taken into account in order to
preserve the $K$-invariance and to recover the correct response
functions.
 
In both the Hamiltonian and Lagrangian descriptions, $K$--invariance
leads to the existence of zero energy excitation modes for the
longitudinal momentum density at finite wavevectors.  However, these
modes do not couple to any physical observables.

We have also investigated the single fermion Green's function and the
fermion effective mass, in the context of our Hamiltonian dipolar
formulation, in the small--$Q$ limit, finding results identical to
those of the conventional FCS calculation for the infra-red
singularities of the effective mass and the single particle Green's
function. These issues were discussed in Secs.  (\ref{sec:smallQ}) and
(\ref{sec:mstar}), where we also discussed the difference between the
small $Q$ model and the physical model with regard to the ultra-violet
renormalization of the effective mass. In this discussion we argued
that present numerical calculations based on trial wave functions are
well suited to study the ultra-violet renormalization but are
inadequate for revealing the infra-red one.  We also showed that the
dipolar approach is free from the inconvenient feature of the
conventional FCS approach that the overlap between the Green's
function for the bare CS fermion and the low-energy quasiparticle
sector vanishes in the limit of an infinite system due to coupling to
the high-energy magnetoplasmons.

The action (\ref{low-e-action}) and the Hamiltonian (\ref{hd}) predict the
same electronic response, despite their very different appearances. As we saw
in Secs. (\ref{sec:Lagrange}) and (\ref{sec:Hamilton}), these differences
originate from a different identification of the position of the
quasi--particle. In Sec.  (\ref{sec:Lagrange}) the position of
quasi--particles is (after averaging over a scale $\sim Q^{-2}$) identical to
the position of the electrons, while in Sec.  (\ref{sec:Hamilton}) the
quasi--particles are shifted halfway from the electrons towards the
correlation holes.  This assignment of a position to a quasi-particle is
arbitrary, particularly since the quasi--particles are dipolar, and  different
assignments should not lead to different predictions regarding the physical
electronic response. The two assignments we used may be argued to be the two
natural ones, but are obviously not the only possible ones.  

An additional difference between the two dipolar descriptions employed
in Secs. (\ref{sec:Lagrange}) and (\ref{sec:Hamilton}) arises due the
time derivative in the last term of the effective low-energy action
Eq.\ (\ref{low-e-action}). When the fields $\newpsidagger,\newpsi$
entering the low-energy action of Eq.\ (\ref{low-e-action}) are
transformed into quantum mechanical operators, their commutation
relations are not canonical. Consequently, the quasi-particle density
operators at different points (or different wave vectors) do not
commute. Preliminary study of these commutation relations indicate
that in the limit of small wavevectors, the commutation relations
between the density operators at different wave vectors is the
commutation relation characterizing the electronic density operator
after projection to the lowest Landau level\cite{unpublished,Shankar}.

Our attempt to proceed from the small $Q$ model to the physical model
of unlimited $Q$ is based on the conjecture that $K$ invariance is not
a property only of the small $Q$ model, but rather a consequence of
gauge invariance, which should hold also in the physical case. This
belief is strongly supported by the analyses of Haldane\cite{Haldane1},
who worked with trial wavefunctions confined to the lowest Landau
level.  However, the approximate unitary transformations (\ref{U})
suggested by MS and used in our Sec.  (\ref{sec:Hamilton}) do not
preserve $K$--invariance when $Q$ is taken to be large, even if the
driving force wave vector $q$ remains very small.  Consequently, we
cannot use this transformation for studying the physical case.
Rather, we showed, by constructing a Fermi liquid theory, that even
without a small $Q$ cut-off, the combination of $K$--invariance and
the dipolar relation (\ref{assum1}), together with conventional Fermi
liquid assumptions, is bound to lead to an electronic density--density
response function of the form (\ref{keeflt}), which is identical in
$q,\omega$ dependence to the small $Q$ result, Eq. (\ref{kee}).  In
particular, the static limit of this response function predicts the
$\nu=1/2$ state to be compressible.

We remark that in the case of the three-dimensional electron gas, the
low energy quasiparticles in the Bohm-Pines formulation\cite{Bohm}
form a more-or-less normal fermi liquid, with nothing similar to the
$K$-invariance and the vanishing of $1+F_1$ which we find in the CS
system.  In the three-dimensional electron system, the fact that the
low-energy quasiparticles are neutral thus leads to incompressibility,
in the sense that the electron density-response function, reducible
with respect to the Coulomb interaction, vanishes $ \propto q^2$.

To conclude, we believe that the dipolar approach to the $\nu=1/2$
problem leads to a better understanding of the low energy excitations
of the $\nu=1/2$ state, and predicts the same electronic density
response as the FCS approach. Whether the dipolar approach can be used
to improve our understanding of topics that were not discussed here,
such as the electronic tunneling density of states, filling factors
away from $\nu=1/2$ and the effect of disorder, is an open question
under extensive study\cite{DHLee,Shankar,unpublished}.

{\bf Acknowledgments:} We are grateful to F.D.M. Haldane, G.  Murthy,
N.  Read and R.  Shankar for helpful discussions.  BIH and AS are
grateful to the Institute of Theoretical Physics at the University of
California in Santa Barbara, where part of this work was carried out.
This research was supported in part by the National Science Foundation
under Grants No. PHY94-07194 and DMR-94-16910 (BIH), by the US-Israel
Binational Science Foundation (95-250) (BIH and AS), by the Minerva
foundation (FvO and AS) by the Israel academy of Science and the V.
Ehrlich career development chair (AS).  FvO thanks the Weizmann
Institute of Science and the Aspen Center for Physics for hospitality
during several extended stays.

\appendix

\section*{The  5 $\times$ 5 Response Matrix $\Pi$}

In this appendix we outline the calculation of the response matrix
$\Pi_{\mu \nu}$ for free fermions in zero magnetic field (See Eq.
\ref{pifive}).  Here $\Pi_{\mu \nu}(\vec q, \omega)$ is the 5 by 5
matrix defined as the Fourier transform of the retarded correlator $i
\theta(t) \langle [\hat O_\mu(t,\vec r), \hat O_\nu(0, \vec r')]
\rangle$ where the operator $\hat O_\mu$ takes the values $\rho, C_l,
g_l, C_t, g_t$ respectively for the five different values of the
indices $\mu$ and $\nu$.  For example, $\Pi_{\rho, C_l}$ is the
Fourier transform of the retarded correlator $i \theta(t)\langle [
\rho(t, \vec r), C_l(0, \vec r')] \rangle$.  Note also that in Eqs.
\ref{pifive} and \ref{vint}, the rows $\mu$ of the response matrix are
arranged in the order $\rho, C_l, g_l, C_t, g_t$.  We can then write
the elements of this correlation function in terms of the matrix
elements of time independent (Schroedinger representation) operators
as
\begin{eqnarray}  \label{eq:PPmat}
  \Pi_{\mu \nu}(\omega, \vec q) &=& \int \frac{d \vec{k}}{(2 \pi)^2}
  [f(\omega_{\vec k + \vec q}) - f(\omega_{\vec k})]  \\
&\times& \frac{\langle \vec k + \vec q | \hat O_\mu(\vec q) |
  \vec k \rangle \, \langle  \vec k | \hat O^*_\nu(\vec q) | \vec k + \vec q
  \rangle  
}{\omega -
  \omega_{\vec k+\vec q} +
  \omega_{\vec k} + i 0^+}     \nonumber
\end{eqnarray}
where $f$ is the Fermi function and $\omega_k = k^2/(2m)$.  The matrix
elements can be calculated directly to give
\begin{eqnarray} 
  \langle \vec k + \frac{\vec q}{2} | \rho(\vec q) |\vec k -
  \frac{\vec q}{2} \rangle &=& 1 \\ 
  \langle \vec k + \frac{\vec q}{2} | \vec g(\vec q) |\vec k -
  \frac{\vec q}{2} \rangle &=&  \vec k \\
    \langle \vec k + \frac{\vec q}{2} | \vec C(\vec q) |\vec k -
  \frac{\vec q}{2}\rangle &=& \nonumber
    i (\omega_{\vec k + \frac{\vec q}{2}} - \omega_{\vec k -
  \frac{\vec q}{2}})  \\ & & \times \langle \vec k + \frac{\vec q}{2} | \label{eq:cinfo}
    \vec g(\vec q) |\vec k - \frac{\vec q}{2}\rangle 
    \\ &\approx& \frac{i \vec q \cdot
  \vec k}{m} \vec k  
\end{eqnarray}

With some difficulty we can now perform the integrals in Eq.
\ref{eq:PPmat} to obtain all of the elements of the matrix $\Pi$.
However, one can save a great deal of effort by relating some of these
elements to response functions that are well known in the literature,
by noting certain relations between the different elements of the
matrix $\Pi$, and by noting that some elements are zero by symmetry.

We begin by examining the symmetry of Eq. \ref{eq:PPmat}.  In
particular, we examine the $\theta$ dependence of the integrand where
$\theta$ is the angle that $\vec k$ makes with the $\bf {\hat x}$
axis.  We note that the integrand can be written as matrix elements
times some rational function of $\cos\theta$ only (not a function of
$\sin\theta$).  The matrix elements of $\rho, g_l$ and $C_l$ are also
analytic functions of $\cos \theta$ whereas the the matrix elements of
$g_y$ and $C_y$ can be written as $\sin \theta$ times an analytic
function of $\cos \theta$.  The integral over $\theta$ in Eq.
\ref{eq:PPmat} is nonzero only if there are an even number of powers
of $\sin \theta$.  Thus, we have $\Pi_{\nu,\mu} = \Pi_{\mu, \nu} = 0$
for $\nu = g_l, \rho$, or $C_l$ and $\mu = g_t$ or $C_t$ which gives
the 12 zero matrix elements shown in Eq. \ref{pifive}.  We note that
these are zero by time reversal symmetry for a system in zero magnetic
field, just as the Hall coefficient is zero.
 
We now focus on elements of the response matrix $\Pi_{\mu \nu}$ where
neither $\mu$ nor $\nu$ is either $C_l$ or $C_t$.  These elements of
$\Pi$ (a 3 by 3 submatrix) are very closely related to the
electromagnetic response of a free 2DEG to a external perturbing
(scalar or vector) potential.  The response matrix $\Kfr$ is defined as
$j_{\mu} = \Kfr_{\mu \nu} A^{ext}_\nu$ where $\mu$ here takes the values
$0, l$ and $t$ ($j_0$ is the density and $A_0$ is the scalar
potential, $j_l$ and $j_t$ are the respective components of the charge
current, and $A_l$ and $A_t$ are the components of the vector
potential).  The response matrix $\Kfr$ has been calculated many times
before (See for example, Ref. \onlinecite{HLR}.  Note that in this
reference $\Kfr_{tt}$ is referred to as $K^0_{11}$).  Using Kubo formula,
or simple linear response, the matrix $\Kfr$ can be related to the retarded 
correlation function of paramagnetic currents plus a
diamagnetic term.  These  correlation functions are
precisely the correlation functions we need for calculation of the 3 by
3 submatrix of $\Pi$.  Also noting that the definition of $\vec g$
differs from the definition of the physical paramagnetic current by a
factor of $m$ (the charge current is $\vec j = (\vec g - e \rho \vec
A)/m$), we then can obtain the relation
\begin{equation}
\Pi_{\mu \nu} = m^{\alpha(\mu) + \alpha(\nu)} \Kfr_{\mu \nu}  + m n_0
\alpha({\mu}) \delta_{\mu \nu}   
\end{equation}
where $\alpha(\mu) = 0$ for $\mu = \rho$ and $\alpha(\mu) = 1$ for
$\mu = g_l$ and for $\mu = g_t$.  Here, the addition of $m n_0$ is the
contribution of the diamagnetic term.

Current conservation and gauge invariance restrict the elements of
the response matrix to have the symmetry $\Kfr_{\mu l} = (\omega/q)\Kfr_{\mu
  0}$ and similarly $\Kfr_{l \mu} = (\omega/q) \Kfr_{0 \mu}$, such that
there are only 3 independent elements of the 3 by 3 matrix $\Kfr$, of
which one is zero by time reversal symmetry.  Thus, we are left with
two independent nonzero elements of $\Kfr$ which are $\Kfr_{00} = \Pi_{\rho
  \rho} $ and $\Kfr_{tt} = \Pi_{g_t g_t}/(m^2) - n_0/m $ (Note in Eqs.
\ref{free-fermion-jja} we refer to $\Pi_{\rho \rho}$ and $\Pi_{g_t
  g_t}$ as $\Pi_{00}$ and $\Pi_{tt}$ respectively).

To calculate the remaining elements of $\Pi$, we first focus on zero
frequency.  Assuming $q$ is small, we can replace the difference of
Fermi functions in Eq. \ref{eq:PPmat} by $(\omega_{k + q} - \omega_k)
df/d\omega$ which then leads to the simplification
\begin{eqnarray} \nonumber
  \Pi_{\mu \nu}(\omega=0, \vec q) &=& \int \frac{d \vec{k}}{(2 \pi)^2}
  \langle \vec k + \vec q | \hat O_\mu(\vec q) |
  \vec k \rangle \\ &\times& \langle  \vec k | \hat O_\nu^*(\vec q) | \vec k + \vec q
  \rangle \delta(\omega_k  - E_f)   \label{eq:PPmat2}
\end{eqnarray}
with $E_f$ the Fermi energy.   Plugging in the above matrix elements
and performing this simple integration then yields the zero 
frequency results
\begin{eqnarray}
   \Pi_{\rho, C_l}(\omega=0, \vec q) &=&  \Pi_{C_l,
   \rho}^*(\omega =0, \vec q)  = i q n_0 + \ldots  \\
   \Pi_{C_l, C_l}(\omega=0, \vec q) &=& 3 \pi (q n_0)^2/m + \ldots  \\
   \Pi_{C_t, C_t}(\omega=0, \vec q) &=&  \pi (q n_0)^2/m + \ldots   
\end{eqnarray}
where the $\ldots$ indicates higher orders in $q$.  Note that all
other nonzero elements of the matrix $\Pi_{\mu, \nu}$ which have
either $\mu$ or $\nu$ equal to $C_l$ or $C_t$ are higher
order in $q$ at zero frequency.

Finally, we need to find the nonzero frequency results for all matrix
elements $\Pi_{\mu \nu}$ with either $\mu$ or $\nu$ equal to $C_l$ or
$C_t$.  To do this, we note that we can use Eq. \ref{eq:cinfo} to
relate the $\vec C$ matrix elements to $\vec g$ matrix elements and
insert these into
Eq. \ref{eq:PPmat}  such that we have, for example, 
\begin{eqnarray}
&  &\Pi_{C_j,\nu}(\omega, \vec q) = \int \frac{d \vec{k}}{(2 \pi)^2}
  [f(\omega_{\vec k + \vec q}) - f(\omega_{\vec k})] \\ & & \times
  \frac{i (\omega_{\vec k + \vec q} - \omega_{\vec k}) 
    \langle \vec k + \vec q | g_j (\vec q) |
  \vec k \rangle \, \langle  \vec k | \hat O^*_\nu(\vec q) | \vec k + \vec q
  \rangle  
}{\omega -
  \omega_{\vec k+\vec q} +
  \omega_{\vec k} + i 0^+}  \nonumber
\end{eqnarray}
where $j$ takes the value $l$ or $t$.  We then rewrite the frequency
factors as
\begin{equation}
\frac{\omega_{\vec k + \vec
    q} - \omega_{\vec k}}{\omega - \omega_{\vec k+\vec q} +
\omega_{\vec k} + i 0^+} = -1 + \frac{\omega}{\omega - \omega_{\vec k+\vec q} +
\omega_{\vec k} + i 0^+}
\end{equation}
at which point we see that we have a frequency independent piece (from
the ``1'') and a frequency dependent piece which is precisely $i
\omega$ times $\Pi_{g_j, \nu}$.  Thus we have
\begin{equation}
  \Pi_{C_j, \nu}(\omega, \vec q) = \Pi_{C_j, \nu}(\omega = 0 , \vec q)
  + i \omega \Pi_{g_j, \nu}(\omega, \vec q)
\end{equation} 
and we can similarly derive
\begin{equation}
  \Pi_{\nu, C_j}(\omega, \vec q) = \Pi_{\nu, C_j}
  -  i \omega \Pi_{\nu, g_j}(\omega, \vec q)
\end{equation}
with $j = l,t$. Using these laws we can then completely calculate the
rest of the matrix $\Pi$ with a minimum of effort.

\iflongversion
\typeout{Skippingmulticols}
\else
\end{multicols}
\fi

\begin{references}
\bibitem{compdef} By finite compressibility, we mean that the static
  density-density linear response function $\chi_{\rho \rho} (q)$
  approaches a finite constant for wave vector $q \to 0$, in the case
  of short-range interactions, or varies like $1/v(q)$ for long-range
  interactions $v(q)$.  The linear response function is defined for
  infinitesimal density fluctuations $\delta \rho^e$, at finite $q$,
  in the limit $T=0$, for a system without impurities.  Its
  applicability to finite density fluctuations is restricted to values
  of $\delta \rho^e$ that are small compared to $q \kf/ 4 \pi$. (This
  assures that the effective cyclotron radius $R_c^*$ in the
  Chern-Simons field produced by $\delta\rho^e$ is large compared to
  $q^{-1}$.)  The order of limits for $\delta \rho^e \to 0$ and $q \to
  0$ must be specified because, at $q=0$, the energy is a singular
  function of $\delta \rho^e$, due to de Haas-van Alphen-type
  oscillations, as discussed in Ref. \onlinecite{HLR} .
\bibitem{SimonReview} A recent discussion of the relation between the
  wavefunction approaches and the dipole picture is given by
  S. H. Simon in Ref. \onlinecite{CFgeneral}.
\end{references}
\end{document}